\documentstyle[eqsecnum,twocolumn,aps]{revtex}
\begin{document}
\draft
\title{Analytic Properties of  Finite-Temperature
Self-Energies}
\author{H. Arthur Weldon}
\address{Department of Physics, West Virginia University, Morgantown,
West Virginia 26506-6315}
\date{December 24, 2001}
\maketitle

\begin{abstract}
The analytic properties in the energy variable $k_{0}$ of
finite-temperature self-energies are investigated. A typical
branch cut results from  $n$  particles being
emitted into the heat bath and $n'$ being
absorbed from the heat bath.  There are three main results:
First, in addition to the branch points at which the  cuts
terminate, there are  also branch points attached to the
cuts along their length. 
Second,   branch points at $k_{0}=\pm k$ are ubiquitous and for
massive particles they are essential singularities. Third, in a
perturbative expansion using free particle propagators
or in a resummed expansion in which the  propagator
pole occurs at a real energy, the self-energy will have a branch point
at the pole location.
\end{abstract}

\section{Introduction}

At non-zero temperature most examinations of the self-energy
have emphasized one-loop results.  For massless gauge
theories  Braaten and Pisarski showed that all  diagrams with
  one loop   (but any number of external
lines)  will produce effects as large as the tree diagrams and must
therefore be resummed \cite{Pisarski,BP}.  In a resummed expansion the
discontinuities of various one-loop and two-loop diagrams have been
computed in QCD in order to predict  processes relevant for quark gluon
plasmas such as dilepton production, real photon
production, and vector meson production.

There have been few investigations of the analytic properties of
finite-temperature self-energies \cite{Evans,BB} and much of the
emphasis has been on the behavior at zero four-momentum \cite{zero}.
 
The development of the questions investigated in this paper and of
the approach is best illustrated by considering how a familiar
zero-temperature calculation changes at non-zero temperature. 

\subsubsection*{T=0 example}

A typical  example arises for  a massive, self-interacting scalar
field with
${\cal H}_{I}=g^{2}\phi^{4}/4!$. One of the two-loop
contributions to the zero-temperature self-energy is
\begin{eqnarray}
\Pi_{F}(K)=-ig^{4}\int{d^{4}P_{1}\over (2\pi)^{4}}
{d^{4}P_{2}\over (2\pi)^{4}}&&\;{1\over
\big(P_{1}^{2}\!-\!m^{2}+i\epsilon\big)}\nonumber\\
\times 
{1\over\big(P_{2}^{2}\!-\!m^{2}+i\epsilon\big)}&&
{1\over \big(P_{3}^{2}\!-\!m^{2}+i\epsilon\big)},\nonumber
\end{eqnarray}
where $P_{3}=K\!-\!P_{1}\!-\!P_{2}$.
 Direct integration of
the energy variables $P_{01}$ and $P_{02}$ gives
\begin{eqnarray}
\Pi_{F}(K)=-g^{2}\int\!{d^{3}p_{1}\over (2\pi)^{2}}
{d^{3}p_{2}\over (2\pi)^{3}} &&\;{1\over
2E_{1}2E_{2}2E_{3}}\label{piT=0}\\
\times\bigg[{1\over
k_{0}\!-\!E_{1}\!-\!E_{2}\!-\!E_{3}\!+\!i\eta}
-&&{1\over
k_{0}\!+\!E_{1}\!+\!E_{2}\!+\!E_{3}\!-\!i\eta}\bigg],
\label{eq1}\nonumber
\end{eqnarray}
with $E_{j}=(p_{j}^{2}+m^{2})^{1/2}$.
The first
denominator produces a branch point in the self-energy at 
$k_{0}=3E(k/3)=(k^{2}+9m^{2})^{1/2}$ and a branch cut along the
positive
$k_{0}$ axis for
$3E(k/3)\le k_{0}\le\infty$. The second
denominator produces a branch cut for $-\infty\le k_{0}\le
-3E(k/3)$.  From the operator point of view these
two contributions arise from inserting a three-particle
intermediate state in the two time-orderings contained in
$\langle 0|T\big(\phi^{3}(x)\phi^{3}(y)\big)|0\rangle$. 
At  higher orders in perturbation theory the  self-energy
continues to be real and analytic in the open interval
$-3E(k/3)< k_{0} < 3E(k/3)$. 

\subsubsection*{The same example at T$\neq$0}

At non-zero temperature there are four real-time
propagators organized into a 
$2\times 2$ matrix $D_{ij}$ with $i,j=1,2$ \cite{LeBellac,Das}.
The proper self-energy becomes a matrix $\Pi_{ij}$.
In the same $g^{2}\phi^{4}/4!$ theory
the two-loop contribution to the time-ordered self-energy
 $\Pi_{11}$ is
\begin{displaymath}
\Pi_{11}(K)=-ig^{4}\!\int{d^{4}P_{1}\over (2\pi)^{4}}
{d^{4}P_{2}\over (2\pi)^{4}}\,
D_{11}(P_{1})D_{11}(P_{2})D_{11}(P_{3}).
\end{displaymath}
The finite-temperature propagators are
\begin{displaymath}
D_{11}(P)={1+f_{BE}\over P^{2}-m^{2}+i\epsilon}
-{f_{BE}\over P^{2}-m^{2}-i\epsilon},
\end{displaymath}
where $f_{BE}=\big[\exp(\beta|p_{0}|)-1\big]^{-1}$ is the
Bose-Einstein function. Performing the integrations over $P_{01}$
and $P_{02}$  leads to an integrand much more complicated than in
Eq. (\ref{piT=0}). The new integrand can be expressed as a linear
combination of eight terms each of which has a different $k_{0}$
dependence in the denominator. The denominators are of the form
$k_{0}\pm E_{1}\pm E_{2}\pm E_{3}+i\eta(\vec{p}_{j})$, in which
all possible sign combinations occur.
In the various denominators  the sign of the infinitesimal imaginary 
parts are momentum-dependent, which makes the calculation 
tedious and the analytic properties obscure.
The complications of the $i\eta$'s is a result of 
 the absolute value bars in $f_{BE}$.

A much simpler analytic structure is enjoyed by the retarded
self-energy $\Pi_{R}$.
 All four $\Pi_{ij}$ can be expressed in terms of $\Pi_{R}$ and
$\Pi_{A}$, where $\Pi_{A}(K)=\Pi_{R}(-K)$. All four propagators
$D_{ij}$ can be expressed in terms of the retarded and advanced
propagators
$D_{R}$ and $D_{A}$ \cite{Aurenche}.   For example, 
$\Pi_{11}(K)$ is can be expressed as
\begin{displaymath}
\Pi_{11}(K)=[1+f]\,\Pi_{R}(K)-f\,\Pi_{A}(K),
\end{displaymath}
where $f=\big[\exp(\beta k_{0})-1\big]^{-1}$ has no absolute value
bars and is an analytic function of $k_{0}$. The retarded
and advanced self-energies have simple analytic structure:
$\Pi_{R}(K)$ is holomorphic for ${\rm Im}(k_{0})>0$ and 
$\Pi_{A}(K)$ is holomorphic for ${\rm Im}(k_{0})<0$.

To compute the retarded self-energy directly without using the
$\Pi_{ij}$,  one can either use the real-time Feynman rules
expressed in terms of $D_{R}$ and $ D_{A}$\cite{Aurenche,EKW} or use
the imaginary-time Feynman rules
\cite{Kapusta} and then analytically continue in energy. For the
above two-loop example in in $g^{2}\phi^{4}/4!$ theory the
result for the retarded self-energy can be expressed as as sum of
eight contributions:
\begin{eqnarray}
\Pi_{R}(K)=\int {d^{3}p_{1}\over (2\pi)^{3}}
{d^{3}p_{2}\over (2\pi)^{3}}
\sum_{A=1}^{8}{f_{A}(\vec{p}_{1},\vec{p}_{2},\vec{p}_{3})
\over k_{0}+i\eta- \psi_{A}},\label{twolooppi}
\end{eqnarray}
where now all the denominators depend on $k_{0}+i\eta$ 
so that $\Pi_{R}(K)$ is manifestly holomorphic in the upper-half of
the complex
$k_{0}$ plane.  Each $\psi_{A}$ is a sum or difference of the
three  particle energies:
\begin{equation}
\psi_{A}=\pm E_{1}\pm E_{2}\pm E_{3},\label{psi}
\end{equation}
where the eight combinations of the $\pm$ signs account for
the eight different $\psi$'s. The physical interpretation of
all these possibilities is standard \cite{AW1}: The energies that
appear in Eq. (\ref{psi}) with a positive sign correspond to
particles emitted into the heat bath; the energies that appear in
 with a negative sign correspond to particles
absorbed from the heat bath. 

\subsubsection*{Approximate dispersion relations that are real}

Using a conventional propagator which has  poles at
$p_{0}=\pm(p^{2}+m^{2})^{1/2}$ may not be a good procedure because  
thermal corrections will shift the pole to a different location.
 The shift in the pole
location is most dramatic in massless theories. For example, in
massless
$g^{2}\phi^{4}/4!$ the one-loop correction will shift the pole
to $p_{0}=\pm(p^{2}+g^{2} T^{2}/24)^{1/2}$. In massless QED and QCD
the one-loop corrections to the fermion and gauge-boson
self-energies shift the locations of the poles to 
 $p_{0}=\pm E(p)$,
where $E(p)$ is  a complicated, real transcendental function of
momentum. The retarded propagator for any real dispersion
relation is
\begin{equation}
D_{R}(P)={1\over (p_{0}\!+\!i\epsilon)^{2} \!-\!E^{2}(p)}.
\label{retprop}\end{equation}
 The analysis in this paper will apply to propagators of the
form $D_{R}$. This propagator is not as complicated as the
hard-thermal-loop
  propagators
$^{*}S(P)$ and $^{*}D_{\mu\nu}(P)$ for fermion and gauge bosons
\cite{Pisarski,LeBellac,Blaizot,Smilga1}.  The hard-thermal-loop
resummed propagators have the structure
\begin{displaymath}
^{*}D_{R}(P)={1\over
(p_{0}\!+\!i\epsilon)^{2}\!-\!p^{2}\!-\, ^{*}\Pi_{R}(p_{0},p)}.
\end{displaymath}
Although $^{*}D_{R}(P)$ has the same poles on the real
axis at $p_{0}=\pm E(p)$ as Eq. (\ref{retprop}), it also has a branch
cut for space-like momentum, $-p\le p_{0}\le p$.  It can be written
as 
\begin{displaymath}
^{*}D_{R}(P)={N(p_{0},p)\over (p_{0}\!+\!i\epsilon)^{2}
\!-\!E^{2}(p)},\end{displaymath}
where the numerator function $N$ contains the branch cut but has no
poles. It seems quite likely that  the self-energies computed with
$^{*}D_{R}$ would contain all the branch points that will be found
using $D_{R}$ and would contain additional branch cuts 
directly related to the branch cut in $N$. However, this diversion
will not be pursued.
 
\subsubsection*{Generalization to complex dispersion relations}

From physical considerations one knows that particle propagation is
damped at finite temperature. 
One-loop calculations are misleading in that
 the
the solution to $E^{2}=p^{2}+m^{2}+\Pi_{R}^{\rm
1\;loop}(E,p)$ is always  a real energy $E(p)$.
  If one
calculates the self-energy to two-loop accuracy then the pole in
the propagator  to two-loop accuracy will be complex. The
location of this complex pole in the retarded propagator will be
denoted by ${\cal E}(p)$:
\begin{displaymath}
{\cal E}(p)=\omega(p)-{i\over 2}\gamma(p),\end{displaymath}
where $\omega$ and $\gamma$ both real and both positive. 
The corresponding propagator in the pole approximation is
\begin{equation}
D_{R}(P)={1\over \big(p_{0}-{\cal E}(p)\big)
\big(p_{0}+{\cal E}^{*}(p)\big)}.\label{retprop2}
\end{equation}
It has no singularities for ${\rm Im} (p_{0})>0$. When  $p_{0}$
is complex but $p$ is real, it satisfies the condition
$D_{R}(p_{0},p)=\big[D_{R}(-p_{0}^{*},p)\big]^{*}$.
 
It is important to emphasize that the damping function 
$\gamma(p)$ cannot just be invented  for
phenomenological purposes, because when the three-momentum is allowed
to become complex,  
$p_{c}=p+ip'$, then ${\cal E}(p_{c})$ must be an analytic function
of the variable $p_{c}$:
\begin{equation}
{\cal E}(p\!+\!ip')=\omega(p,p')-{i\over
2}\gamma(p,p').\end{equation}
In particular $\omega$ and $\gamma$ must satisfy
the  Cauchy-Riemann conditions:
\begin{eqnarray}
{\partial \omega(p,p')\over\partial p}=&&-{1\over 2}\, {\partial
\gamma(p,p')\over\partial p'}\nonumber\\
{\partial \omega(p,p')\over\partial p'}=&& {1\over 2}\,{\partial
\gamma(p,p')\over\partial p}.\nonumber\end{eqnarray}
This guarantees that, in computing  a self-energy correction, the
integration contours for the momentum variables can be distorted into
the complex plane.  Without this property the locus of
non-analyticity for self-energy corrections would be continuous lines
rather than isolated branch points. 

\subsubsection*{Form of the self-energy}

The following analysis will employ propagators of the general form
in Eq. (\ref{retprop2}). Obviously Eq. (\ref{retprop}) can be
considered as a special case.
A particular multi-loop self-energy diagram will have many
branch cuts and can be written as the sum of various integrals, each of
which displays a unique branch cut \cite{AW2}.  The branch cut of the
most general integral results from an intermediate state in which $n$
particles are emitted into the heat bath and $n'$ particles are
absorbed from the heat bath and is of the form
\begin{equation}
 \Pi_{R}(K)=\int\prod_{j=1}^{n-1}
d^{3}p_{j}
\prod_{\ell=1}^{n'}d^{3}q_{\ell}\;\;{f\big(\vec{p},\vec{q}\big)
\over k_{0}-\psi}.
\label{defPi}
\end{equation}
This definition applies in the region  ${\rm Im} (k_{0})>0$,  where
the retarded self-energy is holomorphic. Branch point will
be sought by analytically continuing away from this region. A
particular Feynman diagram will be the sum of several integrals of
this form, involving different values of
$n$ and $n'$ and often different values of the sum $n+n'$.

Although there are $n+n'$ momenta, one of them is 
 determined by momentum
conservation:
\begin{equation}
\vec{k}
=\sum_{j=1}^{n}\vec{p}_{j}+\sum_{\ell=1}^{n'}\vec{q}_{\ell}.
\label{defmom}\end{equation}
The denominator function $\psi$ sums over the energies of the
emitted particles positively and over the energies of the absorbed
particles negatively:
\begin{equation}
\psi=\sum_{j=1}^{n}{\cal E}(\vec{p}_{j})-\sum_{\ell=1}^{n'}
{\cal E}^{*}(\vec{q}_{\ell}).\label{defpsi1}
\end{equation}
The complex energy ${\cal E}$ has a negative imaginary part and so
 $\psi$ automatically has a negative imaginary part. This guarantees
that $\Pi_{R}(K)$ in Eq. (\ref{defPi}) is holomorphic for 
${\rm Im} (k_{0})>0$. 

  The range of
the  function
$\psi$ depends on the values of
$n$ and
 $n'$.  For $n\ge 2$ and $n'\ge 2$ then
$-\infty\le\psi\le\infty$ and so the self-energy  will 
have a branch cut along the entire length of the real $k_{0}$
axis.  For
$n\ge 2$ and
$n'=1$, momentum conservation forces $\psi$ to
be bounded from below but not from above.
For $n=1$ and $n'\ge 2$ momentum conservation forces $\psi$ to be 
bounded from above but not from below.
 For $n=1$ and $n'=1$ the
range of
$\psi$ will be finite.

Most  points at which $k_{0}=\psi$ will not produce
singularities in  $\Pi_{R}$ because the
integration contours can generally be distorted so that the
integration does not pass over the singularity. (Provided that ${\cal
E}(p)$ is analytic in $p$.)  There are two situations which do produce
singularities
\cite{ELOP,IZ}. 
 The
first, called pinch singularities, occur at values of
$k_{0}$ at which two or more singularities of the integrand pinch the 
integration contour from opposite sides. The necessary condition for a
pinch is the simultaneous vanishing of $k_{0}-\psi$ and of the
derivatives of
$\psi$ with respect to the $\vec{p}$ 's and $\vec{q}$ 's.
The sufficient conditions require more detailed study of the
integrand. The second,
called end-point singularities, occurs at values of $k_{0}$ at
which the singularities of the integrand occur at end-points of
the integration region, in this case from 
$\vec{p}$ and $ \vec{q}$ taking on  values $\pm\infty$.

It is perhaps worth emphasizing that it is the location of branch
points that is under investigation and not the value of the 
discontinuity across  the branch cut. For the present purposes it
does  not matter if the discontinuity can be 
grouped as a product of factors from one side
of the cut or the other \cite{cuts}.

\subsubsection*{Applicability to QCD}

In QCD the quark dispersion relations are 
 different from the gluon dispersion relations
 at the one-loop level and certainly at higher loops.
The analysis presented here applies to any self-energy contribution in
which all of the cut propagators have the same
dispersion relation. 
Thus, for quark self-energies it applies to cuts across
intermediate states with all quarks but no gluons. 
Similarly, for gluon self-energies it applies to
cuts across intermediate states that are composed entirely of gluons
or entirely of  quarks.
The inability to treat intermediate states with mixtures of particle
species is obviously a limitation and it will require more work to
overcome. Even at $T=0$, unequal masses are difficult to treat.

\subsubsection*{Sample result}

A simple but interesting example of the results that will be derived
occurs for self-energy diagrams with a three-particle intermediate state.
Three-particle intermediate states in which all three particles are the
same species  occur at  two-loop order in the following cases:
 (a) scalar field self-energy with $\phi^{4}$ interaction as already
discussed; (b) gluon self-energy in QCD with three-gluon intermediate
state; (c) quark  self-energy in QCD with a three-quark intermediate
state. In all these examples $n+n'=3$. One contribution allows
 two particles to be emitted into the heat bath and one particle
to be absorbed from the heat bath ($n=2, n'=1$). Subsequent analysis
will show that there will be three branch points: viz. at 
$k_{0}=\pm k$ and at $k_{0}=\infty$ with branch cuts connecting them.
The branch points at $k_{0}=\pm k$ will be  essential singularities with
behavior $\exp(3m^{2}k/(K^{2} T))$ as
$K^{2}$ approaches zero from the negative region. Here $m$ is the
effective thermal mass from the large-momentum expansion of the
dispersion relation.

If the single-particle energy ${\cal E}(p)$ used to define the loop
expansion is complex, then only the above three branch points occur.
If, however, a real energy $E(p)$ is used then the self-energy will have
a fourth branch point at $k_{0}=E(k)$. In this situation the propagator
which was assumed to have a simple pole at $k_{0}=E(k)$ turns out to
generate a branch point also at $k_{0}=E(k)$. This  ugliness
infects any perturbative expansion built on a real dispersion relation.

\subsubsection*{Organization}

It is assumed throughout that the branch cuts of
thermal self-energies come entirely from the denominators of the
propagator functions and are not affected by the spin of the
particles.

 Sec. II presents three toy examples of functions with branch
cuts that extend from
$-\infty$ to $+\infty$, as this does not occur for
zero-temperature self-energies.  Two of these examples have
additional branch points on the real axis that illustrate features
that will be found in the actual self-energy. 

Sec. III analyzes the branch
points that occur  for the general intermediate state consisting of 
$n$ particles emitted into the heat bath and $n'$ absorbed from the heat
bath. Sec. III.D summarizes the results and may be read independently of
the development sections.

Sec. IV discusses some  implications of the results. 

There are three appendixes. Appendix A and B contain  detailed
proofs that complete the arguments given in Sec. III. 
Appendix C is an explicit one-loop example that displays the
essential singularity at $k_{0}=\pm k$.  Appendix D is an explicit
two-loop example from Wang and Heinz
\cite{Heinz} that shows both the essential singularity at the
light cone and the branch point at the mass shell.

\section{Simple functions with branch points at
$\pm\infty$}

One of the main results of this paper will be that 
self-energies at $T\neq 0$  not only have branch points at
the ends of their branch cuts but also have extra branch
points  not at the ends but  attached to the
cuts.  Although this is unfamiliar from  $T=0$ physics, it is not
very exotic mathematics. This section contains three
toy examples involving one-dimensional integrals that can
be computed exactly.   

Each  example concerns a  function defined by an integral
 of the form
\begin{equation}
F(\omega)=\int_{-\infty}^{\infty}\!dz\;
{1\over\omega-\psi(z)},
\end{equation}
where $\psi(z)$ is a real function when $z$ is real. When ${\rm
Im}\,\omega>0$  the function is holomorphic and defines the retarded 
form of $F(\omega)$. When
${\rm Im}\,\omega<0$ the function is holomorphic and defines the
advanced form of $F(\omega)$.   

In the following examples the function $\psi(z)$ will be real and
 chosen so that $F(\omega)$ have a branch cut running
from $\omega=-\infty$ to $\omega=\infty$, which separates the
two regions of holomorphicity.  Such a cut requires that
$\psi(z)$ takes on all real values. 

The branch points in the three examples can be found by examining
the integrands and are confirmed by 
explicit integration.
 The discontinuity across the
branch cut is pure imaginary
$F(\omega_{r}\!+\!i\epsilon)-F(\omega_{r}\!-\!i\epsilon)=2i{\rm
Im} F(\omega_{r})$, where
\begin{equation}
{\rm Im} F(\omega_{r})=
-\pi \int_{-\infty}^{\infty}\!\!dz\;\delta\,\Big[\omega_{r}-
\psi(z)\Big].\label{disc}
\end{equation}
The discontinuity formula is not the best way to answer
the question  of whether
$F(\omega)$ has any  branch points at finite real values of
$\omega$  that are attached to the branch cut.  

\paragraph*{ Example 1:}  The first example is
\begin{equation}
f(\omega)=\int_{-\infty}^{\infty}\!
dz\,{1\over \omega-\sinh z}.
\end{equation}
For any real value of $\omega$, positive or negative, there is a real
value of $z$ at which the denominator of the integrand vanishes and
this leads to a branch cut along the entire real axis. The
end-points
$z=\pm\infty$ of the integration produce the  branch points  at
$\omega=\pm\infty$.  Explicit integration confirms this:
\begin{equation}
f(\omega)={1\over\sqrt{\omega^{2}\!+\!1}}
\ln\bigg[{\omega\!+\!\sqrt{\omega^{2}\!+\!1} \over
\omega\!-\!\sqrt{\omega^{2}\!+\!1}   }\bigg].\label{f2}
\end{equation}
There are branch points at $\omega=\pm\infty$, where
the argument of the logarithm vanishes. Inspection shows that Eq.
(\ref{f2}) is discontinuous across the real axis.   If
$\omega$ approaches the real axis from above, then the argument of
the logarithm approaches $e^{-i\pi}|R|$; if
$\omega$ approaches the real axis from below, then the argument of the
logarithm approaches
$e^{i\pi}|R|$.  In this example there are no 
branch points at finite values of $\omega$. This is the type of
behavior that is usually thought to be typical of finite
temperature field theory. 

\paragraph*{Example 2:}  Next consider
\begin{equation}g(\omega)=\int_{-\infty}^{\infty}\!\!dz
\;{1\over \omega-z^{3}}.\end{equation}
This integral has end-point singularities
 at $\omega=\pm\infty$ and also
  a pinch singularity at $\omega=0$. 
The pinch occurs  because at $z=0$  both $\psi(z)$ and $d\psi/dz$
vanish \cite{ELOP,IZ}. 

The integral may easily be evaluated by Cauchy's
theorem. For any $\omega$ the  integrand contains three
simple poles as a function of $z$.
When $\omega$ is in the upper half-plane, there are two poles in $z$ 
above the real axis and one pole below. Integration
 gives
\begin{equation} {\rm Im}
\omega>0:\hskip1cm
g(\omega)={2\pi i\over 3}\;{e^{-i2\pi/3}\over
\omega^{2/3}}.\label{ex2}\end{equation} When $\omega$ is in the
lower half-plane, then $g(\omega)$ is the complex conjugate of the
above:
\begin{equation}
{\rm Im}\omega<0:\hskip 1cm  
g(\omega)={-2\pi i\over 3}\;{e^{i2\pi/3}\over
\omega^{2/3}}.\end{equation} As expected, $g(\omega)$  has a
branch cut along the full length of the real axis with branch
points at
$\omega=\pm\infty$. The new feature is the third branch point at
$\omega=0$.  

It is useful to investigate the analytic structure a bit more. Let
$\omega_{0}$ lie in the upper half-plane, where
$g(\omega)$ given by Eq. (\ref{ex2}) is analytic.
To explore $g(\omega)$ in the neighborhood of
$\omega_{0}$, put $\omega=\omega_{0}+re^{i\phi}$ with $r$ real.
As $\phi$ increases from $0$ to $2\pi$, $\omega$ moves in a circle of
radius $r$ centered on $\omega_{0}$. This circle can pass into the lower
half-plane since Eq. (\ref{ex2}) can be analytically continued into the
lower half-plane. If $r<|\omega_{0}|$ the the circle will not pass
around the origin and the function
$(\omega_{0}+re^{i\phi})^{2/3}$ will have the same value at $\phi=0$ and
at $\phi=2\pi$.  However, if $r>|\omega_{0}|$ then $\omega$ 
will encircle the origin and $g(\omega)$ will not return
to its original value. To clarify this, choose $\omega_{0}=0$ so
that 
$\omega=re^{i\phi}$. Then when
$\phi$ increases from $0$ to $2\pi$,
$\omega^{2/3}$ will return to the value $e^{4\pi i/3}\omega^{2/3}$
and $g(\omega)$ will return to the value
\begin{equation}
g_{II}(\omega)={2\pi i\over 3}\;{1\over\omega^{2/3}}.
\end{equation}
This shows that the function $g(\omega)$ has a branch point at
$\omega=0$ in addition to those at $\omega=\pm\infty$.
(If $\omega$ encircles the origin two more times in a counter
clock-wise direction then $g(\omega)$ will return to the original
value in Eq. (\ref{ex2}).)

\paragraph*{Example 3:} 
The third example is
\begin{equation}
h(\omega)=
\int_{-\infty}^{\infty}\!\!dz
\;{1\over 2\omega-z^{3}+3z}.\end{equation}
In addition to end-point
singularities at
$\omega=\pm\infty$, this integral has pinch singularities at
$\omega=\pm 1$. The pinch singularities arise because
$\psi(z)=z^{3}\!-\!3z$ has a local maximum at $z=1$ and a local
minimum at
$z=-1$. Consequently the denominator has a double zero at
$\omega=\pm 1$ \cite{ELOP,IZ}.

Since the integrand has three simple poles, 
the integral can be performed  using Cauchy's theorem. For ${\rm
Im}\,\omega>0$ the result is
\begin{equation}
{\rm Im}\omega>0:\hskip0.2cm h(\omega)={2\pi i\over 3}\;{1\over 
e^{2\pi i/3}A^{2/3}
+e^{-2\pi i/3}B^{2/3}+1},\label{ex3}
\end{equation}
 where
\begin{displaymath}
A=\omega+\sqrt{\omega^{2}\!-\!1}
\hskip1cm B=\omega-\sqrt{\omega^{2}\!-\!1}.
\end{displaymath}
As expected, $h(\omega)$ has four branch points on the real axis:
at $\omega=-\infty$ where A vanishes; at $\omega=\infty$, where B
vanishes; and at $\omega=\pm1$, where A and B have branch points.

The discontinuity of $h(\omega)$ across the real axis can be
computed either directly from Eq. (\ref{ex3}) or by using Eq.
(\ref{disc}). By either method the result is
\begin{eqnarray}
\omega^{2}>1:\hskip0.3cm {\rm Im}\, h(\omega)=&&
-{\pi \over 3}\,{1\over A^{2}+B^{2}+1}\\
\omega^{2}<1:\hskip0.3cm {\rm Im}\, h(\omega)=&&
{2\pi \over 3}\,{1\over 2\cos\big[(2\theta+2\pi)/3\big]+1},
\end{eqnarray}
where for $-1<\omega<1$ the angle $\theta$ is defined by
$\omega=\cos\theta$. The imaginary part has a different value as
$\omega$ approaches 1 from above or from below. For infinitesimal
$\epsilon$, 
\begin{eqnarray}
\omega=1+\epsilon:\hskip0.7cm {\rm Im} h(1+\epsilon)=&&-{\pi\over
9}\\
\omega=1-\epsilon:\hskip0.7cm
{\rm Im} h(1-\epsilon)=&&-\infty.\end{eqnarray}
Thus the imaginary part of $h(\omega)$ is discontinuous at
 at the branch point. This method will be used in Appendix D.

\section{Branch points of self-energies}

This section will examine the general problem of a perturbative
expansion based on propagators of the form Eq.
(\ref{retprop2}) in which ${\cal E}(p)$ is any single-particle
energy,  real or complex.  A summary
of this section is given in III.D.

\subsection{Branch points for $n$ emissions with no absorptions}

The simplest type of branch cuts are those that come from the
production of $n$ particles.  
After integrations over the time-like
components of  the loop momenta, the retarded thermal, self-energy
can be written as an integral over  $n\!-\!1$ independent three
momenta:
\begin{equation}
\Pi_{R}(k_{0},k)=\int\!{ d^{3}p_{1}d^{3}p_{2}\dots
d^{3}p_{n-1}\;\;f(\vec{p}_{j})\over
k_{0}-\psi}.
\label{NpirealE}\end{equation}
The numerator $f(\vec{p}_{j})$ will depend on temperature and 
and on the spins of the particles.  The
denominator function $\psi$ is 
\begin{displaymath}
\psi=\sum_{j=1}^{n}{\cal E}(\vec{p}_{j}).
\end{displaymath}
 The
momentum of the last particle, viz. $\vec{p}_{n}$, is determined by
momentum conservation
\begin{displaymath}
\vec{k}=\sum_{j=1}^{n}\vec{p}_{j}.\end{displaymath}

 A value of $k_{0}$ that makes the denominator of the integrand
in Eq. (\ref{NpirealE}) vanish
will rarely produce a singularity in $\Pi_{R}$ because the integration
contour can be distorted into the complex plane so as to avoid the point at
which the denominator vanishes. Another way to describe this situation is
to focus on the values of $\vec{p}_{j}$ that make the denominator vanish
for a particular $k_{0}$. As
$k_{0}$ varies, the location of the critical $\vec{p}_{j}$ varies. At a
particular
$k_{0}$ the singularity may move onto the real $\vec{p}_{j}$ axis. This
will generally not produce a singularity of the function
$\Pi_{R}$ because the contour can be distorted so as to avoid
 the singularity \cite{ELOP,IZ}.
However if two singularities move so as to pinch the contour
between them at a particular $k_{0}$ then the function $\Pi_{R}$
will have a singularity at that $k_{0}$. The necessary condition
for the denominator of Eq. (\ref{NpirealE}) to have a double pole at
some particular $k_{0}$ requires that both the denominator and its
first derivative vanish
\cite{ELOP,IZ}.

It is convenient to implement momentum conservation by employing
a Lagrange  multiplier $\vec{v}$ and defining a new function
$\Psi$ as
\begin{equation}
\Psi=
\sum_{j=1}^{n}{\cal E}(\vec{p}_{j})
+\vec{v}\cdot\Big(\vec{k}-\sum_{j=1}^{n}\vec{p}_{j}
\Big).
\end{equation}
Any point at which the derivatives of $\Psi$ with respect to
$\vec{p}_{1},\dots,\vec{p}_{n}$ and $\vec{v}$ all vanish, will be a
point at which the derivatives of  $\psi$  with respect to
$\vec{p}_{1},\dots,\vec{p}_{n-1}$ vanish while keeping momentum
conserved.

To proceed further it is helpful to introduce the group velocity 
\begin{equation}
V(p)={d{\cal E}(p)\over dp},\end{equation}
which may be complex when ${\cal E}$ is complex.
 The pinch
conditions 
\begin{displaymath}
0={\partial\Psi\over\partial\vec{p}_{j}}=\hat{p}_{j}V(p_{j})-\vec{v},
\end{displaymath}
imply that all the  $\vec{p}_{j}$
are equal. The common value of
$\vec{p}_{j}$ is determined by extremizing with respect to the Lagrange
multiplier:
\begin{displaymath}
0={\partial\Psi\over\partial\vec{v}}=\vec{k}-\sum_{j=1}^{n}\vec{p}_{j},
\end{displaymath}
and this fixes $\vec{p}_{j}=\vec{k}/N$. The extreme value of
$\psi$  is
\begin{equation}
\psi_{\rm ext}=n\,{\cal E}\big(\vec{k}/n\big).\label{psiextreme}
\end{equation}
Thus there will be branch point at $k_{0}=n{\cal E}(\vec{k}/n)$.
 For the free-particle dispersion relation
the branch point is at
$k_{0}=\sqrt{k^{2}+(nm)^{2}}$.
When ${\cal E}$ is
an effective thermal energy, the branch point at $k_{0}=n\,E(k/n)$ 
 will be temperature-dependent.  In either case, the branch cut
runs parallel to the positive  $k_{0}$ axis and terminates with a
branch point at $k_{0}=\infty$.

For the related situation of $n$ particles absorbed from the thermal
bath, then $\psi=-\sum_{j=1}^{n}{\cal
E}^{*}(\vec{p}_{j})$. This produces a branch point at $k_{0}=-n{\cal
E}^{*}(\vec{k}/n)$ and a branch cut which runs parallel to the negative 
$k_{0}$ axis and terminates at $-\infty$.

\subsection{Branch points that only occur for real group velocities}

The part of the self-energy which has an intermediate state consisting
of  $n$ emitted particles and $n^{\prime}$ absorbed particles has
the form
\begin{equation}
\Pi_{R}(K)=\int\!\prod_{j=i}^{n-1} d^{3}p_{j}
\prod_{\ell=1}^{n'}d^{3}q_{\ell}\;\;{f\big(\vec{p},\vec{q}\big)
\over k_{0}-\psi}
\end{equation}
where the momentum $\vec{p}_{n}$ is determined by momentum
conservation, Eq. (\ref{defmom}), and
$\psi$ is given by Eq. (\ref{defpsi1}). 

 To examine for pinch singularities in  momentum space
subject to the constraint of momentum conservation, it is
again convenient to introduce a Lagrange multiplier
$\vec{v}$ and define a new function
\begin{equation}
\Psi=
\sum_{j=1}^{n}{\cal E}(\vec{p}_{j})
\!-\!\sum_{\ell=1}^{n'}{\cal E}^{*}(\vec{q}_{\ell})+\vec{v}\cdot\Big(
\vec{k}-\sum_{j=1}^{n}\vec{p}_{j}-\sum_{\ell=1}^{n'}
\vec{q}_{\ell}\Big).\label{Bpsi}
\end{equation}
The pinch condition
\begin{displaymath}
0={\partial\Psi\over\partial \vec{p}_{j}}=
\hat{p}_{j}V(p_{j})-\vec{v}
\end{displaymath}
implies that all $\vec{p}_{j}$ are equal.
The condition
\begin{displaymath}
0= {\partial\Psi\over\partial \vec{q}_{\ell}}=
-\hat{q}_{\ell}V^{*}(q_{\ell})-\vec{v}
\end{displaymath}
implies that all the $\vec{q}_{\ell}$ are equal.
Eliminating the Lagrange multiplier $\vec{v}$ in these last two
conditions gives
\begin{equation}
\hat{p}_{j}V(p_{j})=-\hat{q}_{\ell}V^{*}(q_{\ell}).\label{cond12}
\end{equation}
The third condition is
\begin{equation}
0={\partial\Psi\over\partial \vec{v}}=\vec{k}
-\sum_{j=1}^{n}\vec{p}_{j}-\sum_{\ell=1}^{n'}\vec{q}_{\ell}.\label{cond3}
\end{equation}

\paragraph*{Case 1. ${\cal E}$ real:}
When the single particle energy ${\cal E}$ is real, it is denoted by $E$.
The group velocity
$V$ is real.  Eq. (\ref{cond12}) implies first  that
$\hat{p}_{j}=-\hat{q}_{\ell}$ and second that $V(p_{j})=V(q_{\ell})$.
This is solved by
$\vec{p}_{j}=-\vec{q}_{\ell}$. Eq. (\ref{cond3}) can then
be solved for $n\neq n'$:
\begin{displaymath}
\vec{p}_{j}={\vec{k}\over n\!-\!n'}\,;
\hskip1cm \vec{q}_{\ell}={-\vec{k}\over
n\!-\!n'}.
\end{displaymath}
(For $n\neq n'$ there is no solution.)
The value of Eq. (\ref{Bpsi}) at the extremum is
\begin{equation}
\psi_{\rm ext}= [n\!-\!n']\,E\Big(k\big/[n\!-\!n']\Big).
\end{equation}
The necessary conditions for a branch point at $k_{0}=\psi_{\rm ext}$
are thus satisfied. Because this extrememum is a saddle point and not a
local maximum or minimum, the conventional experience does not apply.
To demonstrate that there actually is a pinch of the
integration contour requires more analysis. This analysis is done in
Appendix A and confirms that there is a branch point at
$k_{0}=\psi_{\rm ext}$ and also shows that the branch point has
infinitely many sheets.  

For the free
particle dispersion relation the branch point is at
$k_{0}=\sqrt{k^{2}+((n-n')m)^{2}}$.
If $E$ is a temperature-dependent effective
 energy, then the location of the
branch cut will be temperature-dependent. 
 
The most surprising consequence of this is that when
$n\!-\!n'=1$, there is a branch point at $k_{0}=E(k)$, which is
precisely at the location of the pole in the propagator that was
used to define the perturbative series. 
An example of this phenomena occurs in the self-energy
in $\phi^{4}$ theory.
Wang and Heinz \cite{Heinz} have calculated the imaginary part of 
the two-loop self-energy. Appendix D shows explicitly that the
two-loop self-energy has a branch point at the mass-shell.

When $n\!-\!n'=2$ there is a branch point at $k_{0}=2E(\vec{k}/2)$ that
occurs by cutting $n+n'$ propagators.  For
$n'=0$ this is the two-particle normal threshold
already displayed in Eq. (\ref{psiextreme}). But for $n'\neq 0$ the
branch point occurs in more complicated diagrams than 
in Eq. (\ref{psiextreme}). Similarly, for $n\!-\!n'=7$ the branch point 
at $k_{0}=7E(\vec{k}/7)$ occurs in  diagrams with 
$n+n'\ge 7$. 

\paragraph*{Case 2. ${\cal E}$ complex but $V$ real:} It is possible
to have ${\cal E}(p)=E(p)-i\gamma/2$ but $\gamma$ is a
non-zero constant. The true damping cannot be constant, but the
 constant $\gamma$ approximation is sometimes useful
\cite{Henning}. The group velocity $V=dE/dp$ is real so that the
pinch condition is satisfied at the same momenta $\vec{p}_{j}$ and
$\vec{q}_{\ell}$ as in case 1. The only difference is that the
extremum of Eq. (\ref{Bpsi}) is
\begin{equation}
\psi_{\rm
ext}=[n\!-\!n']\,E\Big(k\big/[n\!-\!n']\Big)-i(n\!+\!n'){\gamma\over
2}.\end{equation}
When $n\!-\!n'=1$ the various branch points at
$k_{0}=E(k)-i(n\!+\!n')\gamma/2$ do not coincide with the
single-particle pole at
$k_{0}=E(k)-i\gamma/2$.
The proof in Appendix A includes this case.

\paragraph*{Case 3. ${\cal E}$ and $V$ complex:} When the
single-particle energy
${\cal E}$ is complex, it is difficult to solve Eqs. (\ref{cond12}) and
(\ref{cond3}). The first equation implies that
$\hat{p}_{j}=\pm\hat{q}_{\ell}$. Let us examine the case 
$\hat{p}_{j}=-\hat{q}_{\ell}$. Then 
\begin{displaymath}
V(p_{j})=V^{*}(q_{\ell}).\end{displaymath}
It is possible to invent
an analytic function $V(p)$ that
satisfies this condition at special momentum. However, the branch
points would then be artifacts of the approximation scheme.  The
exact value of the single-particle pole energy,
${\cal E}_{\rm pole}$, the imaginary part  is negative and   vanishes
at zero momentum and at infinite momentum. Therefore its first
derivative must be negative at small momentum and positive at large
momentum. If $p_{j}$ is small and
$q_{\ell}$ is large, it may be possible for
${\rm Im} V(p_{j})=-{\rm Im} V(q_{\ell})$. Whether the real parts
would satisfy 
${\rm Re} V(p_{j})={\rm Re} V(q_{\ell})$ seems unlikely.

\subsection{Essential singularities at $k_{0}=\pm k$}

The branch points discussed above occur when
$\vec{p}_{j}$ and $\vec{q}_{\ell}$ all have a finite magnitude.
Additional branch points
can result from pinches at infinite values of $\vec{p}_{j}$ and
$\vec{q}_{\ell}$.  To investigate these it is necessary to make
some assumption about the behavior of the dispersion relation
${\cal E}(p)$ at large momenta. It will be assumed that
\begin{equation}
p\to\infty:\hskip0.7cm {\cal E}(p)\to p+{m^{2}\over 2p}+\dots
\label{asympt}\end{equation}
and that the imaginary part of ${\cal E}(p)$ falls faster than
$1/p$. 
This is obviously the correct asymptotic behavior for any
theory that is massive at zero temperature. Theories that are
massless at zero temperature require a resummation to obtain a
sensible dispersion relation ${\cal E}(p)$. In this case the
parameter
$m$ plays acts as an effective thermal mass at large momentum.
The asymptotic behavior of one-loop dispersion relations in massless
gauge theories is well-known \cite{LeBellac,Blaizot,Smilga1}. The
asymptotic behavior  Eq. (\ref{asympt})  applies to spinless
fields, to the spinor field components which have the same
helicity as chirality, and to the vector field components that are
transversely polarized.  It does not apply to the spinor field
components that have the helicity opposite to the chirality nor
to the longitudinally polarized vector bosons. (Both these cases
have asymptotic behavior
${\cal E}(p)\to p+Ap\exp(-p^{2}/m^{2})$.  However, in these two cases
the residue of the pole vanishes at large momentum like 
$\exp(-p^{2}/m^{2})$.) 

 To investigate the branch points that can occur at large momenta, it
is useful to  introduce three Lagrange multiplier vectors:
$\vec{v}_{1},\vec{v}_{2},\vec{P}$ and define
\begin{eqnarray}
\Psi=
\sum_{j=1}^{n}{\cal E}(\vec{p}_{j})
-&&\sum_{\ell=1}^{n'}{\cal E}^{*}(\vec{q}_{\ell})
+\vec{v}_{1}\cdot\Big({1\over
2}\vec{k}+\vec{P}-\sum_{j=1}^{n}\vec{p}_{j}\Big)\nonumber\\
+&&\vec{v}_{2}\cdot\Big({1\over 2}\vec{k}-\vec{P}-\sum_{\ell=1}^{n
'}\vec{q}_{\ell}\Big).\label{Cpsi}
\end{eqnarray}
This is equivalent to Eq. (\ref{Bpsi}) because extremizing with
respect to
$\vec{P}$ sets
$\vec{v}_{1}=\vec{v}_{2}$. 
However we will
compute the extrema of Eq. (\ref{Cpsi}) by computing the
derivatives in a different order. The  pair of conditions
\begin{eqnarray}
0=&&{\partial\Psi\over\partial\vec{p}_{j}}=
V(p_{j})\,\hat{p}_{j}-\vec{v}_{1}\nonumber\\
0=&&{\partial
\Psi\over\partial\vec{v}_{1}}={1\over
2}\vec{k}+\vec{P}-\sum_{j=1}^{n}\vec{p}_{j},\nonumber
\end{eqnarray}
imply  that all $\vec{p}_{j}$ are equal and  that they have
the common value
\begin{equation}
\vec{p}_{j}=({1\over 2}\vec{k}+\vec{P})/n. 
\label{commonp}\end{equation}
The pair of conditions
\begin{eqnarray}
0= &&{\partial\Psi\over\partial \vec{q}_{\ell}}=
-V^{*}(q_{\ell})\,\hat{q}_{\ell}-\vec{v}_{2}\nonumber\\
0=&&{\partial\Psi\over\partial\vec{v}_{2}}={1\over 2}\vec{k}-\vec{P}-
\sum_{\ell=1}^{n'}\vec{q}_{\ell}\nonumber
\end{eqnarray}
imply that all the $\vec{q}_{\ell}$ are equal and have the common
value
\begin{equation}
\vec{q}_{\ell}=({1\over 2}\vec{k}-\vec{P})/n'.\label{commonq}
\end{equation}
As a result, 
\begin{displaymath}
\Psi=n\,{\cal E}\bigg({1\over
n}\big|\vec{P}+\vec{k}/2\big|\bigg)-n'\,
{\cal E}^{*}\bigg({1\over n'}\big|\vec{P}-\vec{k}/2\big|\bigg).
\end{displaymath}

 The condition
$0=\partial\Psi/\partial
\vec{P}$ requires
\begin{displaymath}
{\vec{P}+\vec{k}/2\over
|\vec{P}+\vec{k}/2|}\;V\bigg({1\over n}\Big|\vec{P}+{\vec{k}\over
2}\Big|\bigg)=
{\vec{P}-\vec{k}/2\over
|\vec{P}-\vec{k}/2|}\;V^{*}\bigg({1\over n'}\Big|\vec{P}-{\vec{k}\over
2}\Big|\bigg).
\end{displaymath}
Regardless of the value of the group velocity in this equation,
the two vectors $\vec{P}+\vec{k}/2$ and $\vec{P}-\vec{k}/2$
can only be  proportional to each when
$\vec{P}_{\bot}=0$, where
$\vec{P}=\hat{k}P_{\|}+\vec{P}_{\bot}$. When $\vec{P}_{\bot}=0$ the
vectors  multiplying $V$ and $V^{*}$, respectively,  both are equal
to the unit vector $\hat{k}$. 
Thus the
condition reduces to
\begin{equation}
V\bigg({1\over n}\Big|P_{\|}+{k\over
2}\Big|\bigg)=
V^{*}\bigg({1\over n'}\Big|P_{\|}-{k\over
2}\Big|\bigg).
\end{equation}
Because of the presumed asymptotic behavior in Eq. (\ref{asympt}),
this is satisfied in the limit 
$P_{\|}\to\pm\infty$. 
If $V$ is complex, this is the only possible solution.
If $V$ is real, then in addition to the solution for infinite $P_{\|}$
there is also a finite solution 
when $n\neq n'$, namely
$\vec{P}_{\|}+k/2=nk /(n\!-\!n')$. The finite solutions was 
already treated in Sec. III.B and requires no further discussion.

Thus, regardless of the values of $n$ and $n'$, at
$P_{\|}\to\infty$ the necessary conditions are satisfied for a branch
point. From Eqs. (\ref{commonp}) and (\ref{commonq}) and the fact that
$\vec{P}_{\bot}=0$, the   important region of integration is 
\begin{equation}
\vec{p}_{j}={\hat{k}\over n}({k\over 2}+P_{\|});
\hskip0.8cm 
\vec{q}_{\ell}={\hat{k}\over n}({k\over 2}-P_{\|}).\label{pandq}
\end{equation}
The denominator function is
\begin{displaymath}
\psi=n\,{\cal E}\Big({1\over n}\Big|P_{\|}+{k\over 2}\Big|\Big)
-n'\,{\cal E}^{*}\Big({1\over n'}\Big|P_{\|}-{k\over 2}\Big|\Big).
\end{displaymath}
The asymptotic behavior assumed in Eq.
(\ref{asympt}) implies
\begin{equation}
P_{\|}\to +\infty:\hskip0.3cm \psi\to k+{(nm)^{2}\over 2P_{\|}+k}
-{(n'm)^{2}\over 2P_{\|}-k}+\dots
\end{equation}
Therefore the branch point in $\Pi_{R}(K)$ produced  by
the denominator $k_{0}-\psi$ will occur at $k_{0}=k$. 
The region $P_{\|}\to -\infty$ produces
a branch point at $k_{0}=-k$. As in Sec. III.B the arguments thus
far presented are only necessary conditions for a branch point.
To show sufficiency requires a  more detailed analysis and this
is provided in Appendix B. 

\subsubsection*{Why an essential singularity}

 In Sec. A and B the
branch points were produced by particle momenta that were finite.
Here the branch points at $k_{0}=\pm k$ are produced by momenta that
are infinite and this makes it possible to show that the branch
points are  essential singularities. 

The effect comes from  the statistical factor,
$S$, in the integrand of the self-energy contribution that contains $n$
emitted particles and
$n'$ absorbed particles:
\begin{equation}
S=S_{\rm direct}-\sigma S_{\rm inverse},\label{statistical1}
\end{equation}
where $\sigma=1$  for a boson self-energy and $\sigma=-1$ for a fermion
self-energy. The statistical factors are
\begin{eqnarray}
S_{\rm
direct}=&&\prod_{j=1}^{n}[1+\sigma_{j}N_{j}]\,\prod_{\ell=1}^{n'}
N^{*}_{\ell}
\\
S_{\rm
inverse}=&&\prod_{j=1}^{n}N_{j}\,\prod_{\ell=1}^{n'}
[1+\sigma_{\ell}N^{*}_{\ell}]
\end{eqnarray}
where  $\sigma=\pm 1$ for bosons and fermions and
\begin{eqnarray}
N_{j}=&&1/[\exp(\beta {\cal E}(\vec{p}_{j}))-\sigma_{j}]\nonumber\\
N^{*}_{\ell}=&&1/[\exp(\beta
{\cal E}^{*}(\vec{q}_{\ell}))-\sigma_{\ell}]\nonumber.
\end{eqnarray}
Because of Eq. (\ref{asympt}) and (\ref{pandq}), in the region
$P_{\|}\to\infty$  the statistical factor  becomes
\begin{equation}
 S\to \big( e^{\beta k/2}-\sigma e^{-\beta k/2}\big)\,
\exp\big(-\beta P_{\|}\big).\label{statistical2}
\end{equation}

By finding the way in which $P_{\|}$ approaches infinity
as $k_{0}\to k$, one can be more specific about the nature of
the branch point.  Near the branch point,  $k_{0}-k$ is very small but
non-zero, and the self-energy denominator $k_{0}-\psi$ will vanish when
$P_{\|}$ is very large but not actually  infinite. The condition
$k_{0}-\psi=0$ gives a quadratic equation for $P_{\|}$. The two roots are
\begin{eqnarray}
P_{\|\pm}=&&{1\over 4(k_{0}-k)}
\bigg[(n^{2}\!-\!n'^{2})m^{2}\nonumber\\
\pm&&\Big(\big[2k(k_{0}\!-\!k)
-(n^{2}\!+\!n'^{2})m^{2}]^{2}
-[2nn'm^{2}]^{2}\Big)^{1/2}\bigg].\nonumber
\end{eqnarray}
There are two cases to be distinguished.

\paragraph*{Case 1. $n=n'$:} If $k_{0}$ is real and approaches $k$
from below, the root that approaches
$+\infty$ is
\begin{equation}
P_{\|-}={k\over 2}\sqrt{1-{2(nm)^{2}\over k(k_{0}-k)}}.
\label{root}\end{equation}
The  statistical factor
Eq. (\ref{statistical2}) becomes
\begin{equation}
S\to \big( e^{\beta k/2}-\sigma e^{-\beta k/2}\big)
\exp\Bigg(\!-{\beta k\over 2}\sqrt{1-{2(nm)^{2}\over
k(k_{0}-k)}}\;\Bigg).\label{statistical3}
\end{equation}
This is an essential singularity at $k_{0}=k$. If $k_{0}$ is real
and approaches $k$ from below, then $S\to 0$. 
More generally the behavior depends on how $k_{0}-k$ approaches
zero in the complex plane. Appendix C provides a one-loop
calculation with
$n=n'=1$  that displays this behavior in Eq. (\ref{essential1}).

\paragraph*{Case 2. $n\neq n'$:} For definiteness take $n>n'$. Then if 
$k_{0}$ is real and approaches $k$ from above, the root that approaches
$+\infty$ is
\begin{displaymath}
P_{\|+}\to{(n^{2}-n'^{2})m^{2}\over 2(k_{0}-k)}.
\end{displaymath}
The statistical factor  Eq. (\ref{statistical2}) becomes
\begin{equation}
S\to \big( e^{\beta k/2}\!-\!\sigma e^{-\beta k/2}\big)
\exp\Bigg(\!-{\beta(n^{2}-n'^{2})m^{2}\over 2(k_{0}-k)}\Bigg).
\label{statistical4} \end{equation}
This, again, is an essential singularity at $k_{0}=k$, whose
behavior naturally depends upon the direction from which $k_{0}$
approaches $k$.
Appendix D provides a two-loop
calculation with $n=2$, $n'=1$, and $k=0$. The exponent is
predicted to be $-\beta 3m^{2}/2k_{0}$, and this is just what is
found in  Eq. (\ref{essential2}).

\paragraph*{Comment on Hard Thermal Loops:} The one-loop gluon
self-energy has branch points at $k_{0}=\pm k$
\cite{LeBellac,Blaizot,Smilga1}. These come from intermediate states
with
$n=1, n'=1$ (either  two gluons or two fermions). The one-loop
calculations are done using a massless dispersion relation for the
intermediate particles. Therefore $m=0$ in Eq. (\ref{asympt}) so that
$P_{\|+}=k/2$ in Eq. (\ref{root}). Since this momenta is finite, the
statistical factor
$S$ is unremarkable and cannot produce an essential singularity.
Explicit calculations show that the branch points at $k_{0}=\pm k$ are
 logarithmic for the hard thermal loops. 

\subsection{Summary}

In the following summary the single-particle energies ${\cal
E}(p)$ can be complex or real. For results that only apply when  the
energies are real, the block $E$ will be used instead of the
script ${\cal E}$. Real $E$ produces  the exceptional
 branch points discussed in
Sec. III.B and they will  be described in parentheses in the
summary below. The ubiquitous branch points  at
$k_{0}=\pm k$ are always essential singularities and this will not be
repeated each time. 
As noted following Eq. (\ref{defpsi1}),  when  $n$ and/or
$n'$ have the value $1$, the range of
$\psi$ is constrained by momentum conservation and  this often
determines the end points of the branch cuts.  

\subsubsection{Organized by $n'$,  the number of absorptions}

The most concise way to summarize the previous results is
by  $n'$, the number of particles absorbed from the heat bath.

\paragraph*{(a) No absorptions: $n'=0$.} This is the
simplest case and directly analogous to zero temperature.
The branch cut is semi-infinite: $ n\,{\cal E}(k/n)\le
k_{0}\le\infty$.

\paragraph*{(b) One absorption: $n'=1$.} There are two
subcases. If  $n=1$ then
the branch cut is only for space-like four momenta: $-k\le
k_{0}\le k$. If 
$n\ge 2$ then the branch cut is semi-infinite: $-k\le k_{0}
\le\infty$; and there  is a branch point at
$k_{0}=k$. (If the single-particle energies are real, there is an
exceptional branch point at
$k_{0}=[n-1]E(k/[n-1])$. For
$n=2$ the last branch point coincides with the free particle pole at
$k_{0}=E(k)$.)

\paragraph*{(c) Two or more absorptions: $n'\ge 2$.}
There are three subcases. If $n=0$ the branch cut is semi-infinite:
$-\infty\le k_{0}\le -n'{\cal E}^{*}(k/n')$.
If $n=1$ the branch cut is also semi-infinte: $-\infty\le k_{0}\le
k$, and there is an additional branch point at $k_{0}=-k$.
If $n\ge 2$ the
branch cut runs the full length of the real axis:
$-\infty\le k_{0}\le\infty$;  and there are two
additional branch points at  $k_{0}=\pm k$. (If the single-particle
energies are real, there are exceptional branch points for $n\neq n'$ at
$k_{0}=[n-n']E(k/[n-n'])$. Whenever $n-n'=\pm 1$ this last
branch point coincides with the free particle poles at
$k_{0}=\pm E(k)$.)

\subsubsection{Organized by $n+n'$, the number of particles in the
intermediate state}

A particular diagram can generally be cut in several
possible ways.  Each cut is through a particular
number of propagators or equivalently through an
intermediate state with a  particular number of particles.
As a practical matter, this is perhaps the most useful way
to summarize the cut structure.

\paragraph*{(a) Two-particle states:} Two-particle intermediate
states, $n+n'=2$, are possible with a cubic coupling but not with a
quartic coupling.  There are three types of branch cuts. For
$n=2, n'=0$ there will be a semi-infinite cut
$2{\cal E}(k/2)\le k_{0}\le\infty$.  For $n=n'=1$ there will be
a finite length branch cut $-k\le k_{0}\le k$.
For $n=0,n'=2$ there will be a semi-infinite branch cut
$-\infty\le k_{0}\le -2{\cal E}^{*}(k/2)$.

\paragraph*{(b) Three-particle states.}
The possibility $n+n'=3$ occurs with both cubic and quartic coupling. 
There are four types of branch cuts.  For
$n=3, n'=0$ there will be a semi-infinite branch cut
$3{\cal E}(k/3)\le k_{0}\le\infty$. For $n=2, n'=1$ there will be a
semi-infinite branch cut for $-k\le k_{0}\le\infty$ and in addition there
will be a branch point attached to the cut  at $k_{0}=k$. (If the
single-particle energies are real, there will be an exceptional branch
point at 
$k_{0}=E(k)$.) 
 For $n=1$, $n'=2$ the range of the branch cut
is $-\infty\le k_{0}\le k$ with an attached branch point at
$k_{0}=-k$. (If the
single-particle energies are real, there will be an exceptional branch
point at 
$k_{0}=-E(k)$.)  For $n=0$, $n'=3$ the
extent of the branch cut will be $-\infty\le k_{0}\le
-3{\cal E}^{*}(k/3)$.

\paragraph*{(c) Four-particle  states:}
For a cut through $n+n'=4$ propagators, there are
five different types of branch cuts. For $n=4$, $n'=0$
there is only a  four-particle production cut
for $4{\cal E}(k/4)\le k_{0}\le \infty$. For $n=3$, $n'=1$ the
range of the branch cut is $-k\le k_{0}\le\infty$ with
an additional branch point at $k_{0}=k$. (If the
single-particle energies are real, there will be an exceptional branch
point at 
$k_{0}=2E(k/2)$.) For $n=n'=2$ the branch cut runs the full
length of the real axis, $-\infty\le k_{0}\le \infty$ with
two additional branch points at $k_{0}=-k$ and $k_{0}=k$. 
For $n=1$, $n'=3$ the range of the branch cut is
$-\infty\le k_{0}\le k$ with an additional branch point
at $k_{0}=-k$. (If the
single-particle energies are real, there will be an exceptional branch
point at $k_{0}=-2E(k/2)$.)  For $n=0$, $n'=4$
there is only a four-particle absorption cut for
$-\infty\le k_{0}\le -4{\cal E}^{*}(k/4)$.

\section{Comments}

\subsection{Expectations for the exact self-energy}

Section III.A showed that there will a branch point in the retarded
self-energy $\Pi_{R}(k_{0},\vec{k})$ at $k_{0}=n{\cal E}(\vec{k}/n)$
that results from the emission of
$n$ particles and, likewise, a branch point at $k_{0}=-n{\cal
E}^{*}(\vec{k}/n)$ that results from the absorption of $n$ particles
from the heat bath. The effective single-particle energies will
generally be temperature-dependent and complex and so the location of
the branch points will generally be temperature-dependent and complex.
Furthermore, the location of the branch points is model-dependent
in the sense that
one can change the single-particle energies ${\cal
E}$ and thus change the location of the branch points in the
perturbative expansion. 

However the location of the branch points in the exact self-energy
cannot be model-dependent. If one summed the perturbative
self-energy contributions to all orders, the 
 model-dependence of the branch points would disappear
just as the model-dependence of the propagator pole would disappear. 
If the exact propagator has a pole at ${\cal E}_{\rm exact}(p)$,
the exact self-energy should have normal-threshold branch points
at $k_{0}=n\,{\cal E}_{\rm exact}(\vec{k}/n)$ and at $k_{0}=-n\,{\cal
E}_{\rm exact}^{*}(\vec{k}/n)$. In addition, there will be 
essential singularities at $k_{0}=\pm k$. 

\subsection{Bad features of real dispersion relations}

Performing perturbative calculations using the free thermal
propagator or indeed any thermal propagator containing a real
dispersion relation $E(p)$ leads to a self-energy that has branch
points at the perturbative mass-shell, $k_{0}=E(k)$.
 If $\Pi_{R}(k_{0})$ is the retarded
self-energy computed beyond one-loop order using a real energy $E(p)$
to define the perturbation series, then to improve on the value of
$E(p)$ one needs  to solve perturbatively for ${\cal E}_{\rm pole}$
\begin{equation}
{\cal E}_{\rm pole}=\sqrt{E^{2}+\Pi_{R}({\cal E}_{\rm pole})}.
\label{pole1}
\end{equation} 
A perturbative  expansion means that $\Pi_{R}(E)$ and its derivatives
at $E$ are small compared to $E$.  Thus the lowest-order
contribution to the damping rate should be
\begin{equation}
{\rm Im}\,{\cal E}_{\rm pole}\approx {\rm Im} \big(\Pi_{R}(E)\big)/2E.
\end{equation}
However this will fail at two-loop order because $\Pi_{R}(k_{0})$ has
a branch point precisely at $k_{0}=E$. This was first encountered
in calculations of the fermion damping rate in QCD, where it was
found that even with a magnetic mass to eliminate the infrared
divergence there is a branch point at $k_{0}=E$ that
comes from the intermediate state with $n=2, n'=1$ \cite{damping}.
Appendix D contains an explicit calculation in
$\phi^{4}$ theory that shows that the self-energy is not
differentiable at the mass-shell.

If one pretends that the self-energy is 
differentiable at $E$ then the solution to Eq. (\ref{pole1}) would be
\begin{displaymath}
{\cal E}_{\rm pole}=E
+\sum_{s=1}^{\infty}{(-1)^{s}\over s!}\big[\Pi_{R}(E)\big]^{s}
\Bigg[{d^{s-1}\over dk_{0}^{s-1}}{1\over \big[f(k_{0})\big]^{s}}
\Bigg]_{k_{0}=0},
\end{displaymath} 
where $f(k_{0})$ is the function
\begin{equation}
f(k_{0})=-k_{0}-2E
+\sum_{\ell=1}^{\infty}{(k_{0})^{\ell-1}\over \ell !}
{d^{\ell}\Pi_{R}(E)\over dE^{\ell}}.
\end{equation}
However $d^{\ell}\Pi_{R}(E)/dE^{\ell}$ does not exist and thus the
perturbative calculation of ${\cal E}_{\rm pole}$ fails.

There is another consequence of real dispersion relations 
that is curious, though perhaps not as dire. Diagrams in which
two particles are emitted and none absorbed ($n=2, n'=0$) will have
the usual normal threshold branch point at $k_{0}=2E(\vec{k}/2)$. 
All contributions in which $n'\!+\!2$ particles are emitted and $n'$
are absorbed will also have a branch point at $k_{0}=2E(\vec{k}/2)$.
Similarly,  All contributions in which $n'\!+\!3$ particles are
emitted and $n'$ are absorbed will also have a branch point at
$k_{0}=3E(\vec{k}/3)$. These coincidences will be absent if 
a complex dispersion relation is employed. 

\subsection{Good feature of complex dispersion relations}

Any complex dispersion
relation (even if the group velocity is real) will generate a
perturbative expansion that does not have a branch point in the
higher order self-energy at 
$k_{0}={\cal E}(k)$ as show in Sec. III.B.
Let  $\Pi^{\rm eff}_{R}(k_{0})$ be the self-energy computed beyond
one-loop order using the free retarded propagator in Eq.
(\ref{retprop2}). 
 The radiatively-corrected retarded
propagator is
\begin{equation}
D^{\rm eff}_{R}(k_{0})={1\over \big(k_{0}-{\cal E}\big)
\big(k_{0}+{\cal E}^{*}\big)-\Pi^{\rm eff}_{R}(k_{0})}.
\end{equation}
The  pole in this propagator satisfies
\begin{equation}
\big({\cal E}_{\rm pole}-{\cal E}\big)
\big({\cal E}_{\rm pole}+{\cal E}^{*}\big)
=\Pi^{\rm eff}_{R}({\cal E}_{\rm pole}).\label{pole4}
\end{equation}
Since the self-energy $\Pi^{\rm eff}_{R}(k_{0})$ does not have
a branch point at
$k_{0}={\cal E}$, it is infinitely differentiable there.
The perturbative solution to Eq. (\ref{pole4}) is
\begin{displaymath}
{\cal E}_{\rm pole}= {\cal E}
+\sum_{s=1}^{\infty}{(-1)^{s}\over
s!}\big[\Pi^{\rm eff}_{R}({\cal E})\big]^{s}
\Bigg[{d^{s-1}\over dk_{0}^{s-1}}{1\over \big[g(k_{0})\big]^{s}}
\Bigg]_{k_{0}=0},
\end{displaymath}
where $g(k_{0})$ is the function
\begin{equation}
g(k_{0})=-k_{0}\!-\!2\omega
+\sum_{\ell=1}^{\infty}{(k_{0})^{\ell-1}\over \ell !}
{d^{\ell}\Pi^{\rm eff}_{R}({\cal E})\over d{\cal E}^{\ell}}.
\end{equation}
In this case, $g(k_{0})$ does exist and the perturbative
expansion for ${\cal E}_{\rm pole}$ is valid.

\acknowledgments

This work was supported in part by National Science Foundation grant
PHY-0099380.
 
\appendix

\section{Detailed proof of branch point at
\lowercase{$k_{0}=[n\!-\!n']}E(\lowercase{\vec{k}/[n\!-\!n'])$}
for
\lowercase{$n\neq  n'$}}

In Sec. III.B it was shown that if the single-particle energy $E$ is
real, the necessary conditions for a singularity are satisfied  at
\begin{equation}
k_{0}=(n\!-\!n')\, E\big(\vec{k}/[n\!-\!n']\big).
\end{equation}
This Appendix proves sufficiency, viz. that there really is a branch
point.  The momenta which trap the integration contour are near a saddle
point and this makes the analysis different than at zero
temperature.

\subsection{Taylor series expansion of $\psi$ near the
saddle point}

It is convenient to label all the loop momenta as $\vec{p}_{j}$ so
that the denominator function is
\begin{equation}
\psi=\sum_{j=1}^{n}E(\vec{p}_{j})-
\sum_{j=n+1}^{n+n'}E(\vec{p}_{j})\label{Apsi1}.
\end{equation}
Define $\vec{s}=\vec{k}/(n-n')$ and put
  \begin{equation}
 \vec{p}_{j}=\cases{+\vec{s}+\vec{\alpha}_{j},
   & $1\le j\le n$\cr
   -\vec{s}+\vec{\alpha}_{j}, & $n+1\le j\le n+n'$}.
   \end{equation} 
The stationary point of $\psi$ that was found in Sec. III.B  occurs when
all $\alpha_{j}=0$. 
 Momentum conservation requires that
\begin{equation}
0=\sum_{j=1}^{n+n'}\vec{\alpha}_{j}.\label{Amomentum}
\end{equation}

A typical energy can be expanded in a Taylor series to second order
in the small quantities $\vec{\alpha}_{j}$. In
doing this it is convenient to decompose the vectors into
components parallel and perpendicular to $\vec{k}$ or
equivalently to $\vec{s}$. Thus
$\vec{\alpha}_{j}=\hat{k}\alpha_{j\|}+\vec{\alpha}_{j\bot}$.
The Taylor series can then be written to second order as
\begin{displaymath}
E(\vec{s}+\vec{\alpha}_{j})=E(\vec{s})+{dE\over ds}
\alpha_{j\|}
+{1\over
2}A_{\|}\alpha_{j\|}^{2}+{1\over
2}A_{\bot}\vec{\alpha}_{j\bot
}^{2}+\dots\end{displaymath} where
\begin{mathletters}\begin{eqnarray}
A_{\|}= &&{d^{2}E\over ds^{2}} \\
A_{\bot}=&& {1\over s}\,{dE\over ds}.
\end{eqnarray}\label{defA}\end{mathletters}
For a free particle, $E(s)=(s^{2}+m^{2})^{1/2}$, and
$A_{\|}=m^{2}/E^{3}$ and $A_{\bot}=1/E$.
For a quasiparticle dispersion relation, one or
both of $A_{\|}$ and
$A_{\bot}$ could be negative. This will not alter the
following argument. 

When this expansion is inserted into Eq. (\ref{Apsi1}), the
terms linear in
$\vec{\alpha}_{j}$  cancel because of momentum conservation and
leave
\begin{eqnarray}
\psi=(n\!-\!n')E(s)+&&{1\over
2}A_{\|}\bigg(\sum_{j=1}^{n}\alpha_{j\|}^{2}
-\sum_{j=n+1}^{n+n'}\alpha_{j\|}^{2}\bigg)\nonumber\\
+&&{1\over
2}A_{\bot}\bigg(\sum_{j=1}^{n}\vec{\alpha}_{j\bot}^{2}-
\sum_{j=n+1}^{n+n'}\vec{\alpha}_{j\bot}^{2}\bigg)
\label{Apsi2}.
\end{eqnarray}
The constraint in Eq.
(\ref{Amomentum}) means that there are only $n+n'-1$
linearly independent momentum vectors. One
can eliminate the last momentum,  $\vec{\alpha}_{n+n'}$, by
expressing it as the sum of all the other
$\vec{\alpha}$'s.
When this is done, Eq. (\ref{Apsi2}) is no longer  diagonal, 
but it will be real and symmetric and can therefore be
diagonalized by a real rotation. The rotation that diagonalizes the 
terms proportional to $A_{\|}$ will also diagonalize the terms
proportional to $A_{\bot}$.

\subsection{Diagonalization of $\psi$}

 The simplest case is $n=2, n'=1$, i.e. a
three particle intermediate state in which two particles are
emitted and one is absorbed. From the constraint
$\vec{\alpha}_{1}+\vec{\alpha}_{2}+\vec{\alpha}_{3}=0$,
express $\vec{\alpha}_{3}$ in terms of the other two. Then
\begin{eqnarray}
\psi=E(k)+ &&{1\over 2}A_{\|}\big(\alpha_{1\|}^{2}+
\alpha_{2\|}^{2}
-(\alpha_{1\|}+\alpha_{2\|})^{2}\big)
\nonumber\\
+&&{1\over 2}A_{\bot}\big(\vec{\alpha}_{1\bot}^{2}
+\vec{\alpha}_{2\bot}^{2}-(\vec{\alpha}_{1\bot}
+\vec{\alpha}_{2\bot})^{2}\big)\nonumber\\
= E(k)- && A_{\|}\alpha_{1\|}\alpha_{2\|}
-A_{\bot}\vec{\alpha}_{1\bot}\cdot\vec{\alpha}_{2\bot}.
\nonumber
\end{eqnarray} 
This can be easily diagonalized by defining new momenta
$\vec{u}=(\vec{\alpha}_{1}+\vec{\alpha}_{2})/2$
and $\vec{v}=(\vec{\alpha}_{1}-\vec{\alpha}_{2})/2$
so that
\begin{displaymath}
\psi=E(k)+A_{\|}\big(-u_{\|}^{2}+v_{\|}^{2}\big)
+A_{\bot}\big(-u_{\bot}^{2}+v_{\bot}^{2}\big).
\end{displaymath}
The diagonal form is a function of the six Cartesian
components of $\vec{u}$ and $\vec{v}$.  The
diagonal form is traceless with three positive terms
and three negative terms regardless of the signs of
$A_{\|}$ and
$A_{\bot}$.

 For general $n$ and $n'$ express
$\vec{\alpha}_{n+n'}$ in terms of the other $\vec{\alpha}'s$ using Eq.
(\ref{Amomentum}). Then
$\psi$ has the form
\begin{eqnarray}
\psi=(n\!-\!n')E(s)- &&A_{\|}\sum_{i,j=1}^{n+n'-1}
M_{ij}\;\alpha_{i\|}\alpha_{j\|}\nonumber\\
-&&A_{\bot}\sum_{i,j=1}^{n+n'-1}M_{ij}\;
\vec{\alpha}_{i\bot}\cdot\vec{\alpha}_{j\bot},
\end{eqnarray}
where $M_{ij}$ is a simple  numerical matrix. The
eigenvalues of this matrix will be called $\lambda_{j}$.
The matrix can be diagonalized by by a real rotation to a
new basis set
$\vec{\beta}_{j}$. In the new basis $\psi$ will be diagonal
\begin{eqnarray}
\psi=(n\!-\!n')E(s)- &&A_{\|}\sum_{j=1}^{n+n'-1}
\lambda_{j}\;\beta_{j\|}^{2}\nonumber\\
- &&A_{\bot}\sum_{j=1}^{n+n'-1}\lambda_{j}\;
\vec{\beta}_{j\bot}^{2}
\end{eqnarray} 
None of the eigenvalues $\lambda_{j}$ vanish as the
following analysis will show. 

\paragraph*{Case 1. $n'=1$:} When there is  only one absorption
$n'=1$ and any
number of emissions $n\ge 2$, the matrix elements
$M_{ij}$ have the following values
\begin{equation}
M_{ij}=\cases{ 1, & $i\neq j$\cr
0, & $i=j$.}
\end{equation}
There are $n$  eigenvalues of this matrix: 
\begin{equation}
\lambda=\cases{n-1, & degeneracy $=1$ \cr
-1, & degeneracy $= n-1$.}
\end{equation}
Thus the diagonal form of $\psi$ is
\begin{eqnarray}
\psi=(n\!-\!n')E(s)- &&A_{\|}
\bigg((n\!-\!1)\beta_{\|}^{2}-
\sum_{j=1}^{n-1}\beta_{j\|}^{2}\bigg)\nonumber\\
-&&
A_{\bot}\bigg((n\!-\!1)\vec{\beta}_{\bot}^{2}
-\sum_{j=1}^{n-1}\vec{\beta}_{j\bot}^{2}\bigg).
\end{eqnarray}

\paragraph*{Case 2. $n'\ge 2$:} When there are $n'\ge 2$
absorptions and a larger number $n>n'$ of emissions, the matrix
elements are
\begin{equation}
M_{ij}=\cases{ 1, & $i\neq j$\cr
0, & $i=j\leq n$\cr
2, & $i=j\ge n+1$.}
\end{equation}
The matrix  has two
non-trivial eigenvalues:
\begin{displaymath}
\lambda_{\pm}={1\over
2}\bigg[n+n'-1\pm\sqrt{(n+n'-1)^{2}-4(n-n')}\bigg].
\end{displaymath}
Both $\lambda_{+}$ and $\lambda_{-}$ are positive. 
The complete set of  eigenvalues are
\begin{displaymath}
\lambda=\cases{\lambda_{+}, & degeneracy $=1$\cr
\lambda_{-}, & degeneracy $=1$\cr
1, & degeneracy $=n'-2$\cr
-1, & degeneracy $=n-1$.}
\end{displaymath}
Thus $n'$ eigenvalues are positive and $n-1$ are negative.

Thus  there are always $n'$ positive eigenvalues, which will be
labeled
$|\lambda_{1}|,\dots, |\lambda_{n'}|$,  and
 $n-1$ negative eigenvalues, which will be labeled
$-|\lambda_{n'+1}|,\dots ,-|\lambda_{n'+n-1}|$.  Then
$\psi$ has the form
\begin{eqnarray}
\psi=(n\!-\!n')E(s) &&-\sum_{j=1}^{n'}
|\lambda_{j}|\;\Big(A_{\|}\beta_{j\|}^{2}
+A_{\bot}\vec{\beta}_{j\bot}^{2}\Big)\nonumber\\
+ &&\sum_{j=n'+1}^{n+n'-1}|\lambda_{j}|\;
\Big(A_{\|}\beta_{j\|}^{2}
+A_{\bot}\vec{\beta}_{j\bot}^{2}\Big).\label{Apsi}
\end{eqnarray} 
The contribution to the retarded self-energy of this small
region of momentum space is
\begin{equation}
\Pi_{R}(k_{0})=\int\Big(\prod_{j=1}^{n+n'-1}\!d^{3}\beta_{j}
\Big)\;{f(\vec{\beta}_{j})\over k_{0}-\psi}
\end{equation}
In the quadratic approximation, $\psi$ does not depend
separately on all the vectors $\vec{\beta}_{j}$ but only
on two real variables $u$ and $v$ such that
\begin{displaymath}
\psi=(n\!-\!n')E(s)-u^{2}+v^{2}.
\end{displaymath}
As noted earlier, $A_{\|}$ and $A_{\bot}$ are
expected to be positive. If that is the case, 
define $u$ and $v$ by
\begin{eqnarray}
u=&&\bigg[\sum_{j=1}^{n'}
|\lambda_{j}|\;\big(A_{\|}\beta_{j\|}^{2}
+A_{\bot}\vec{\beta}_{j\bot}^{2}\big)\bigg]^{1/2}\\
v=&&\bigg[\sum_{j=n'+1}^{n'+n-1}
|\lambda_{j}|\;\big(A_{\|}\beta_{j\|}^{2}
+A_{\bot}\vec{\beta}_{j\bot}^{2}\big)\bigg]^{1/2}.
\end{eqnarray}
However, if one or both of $A_{\|}$ and $A_{\bot}$ are
negative, then define $-u^{2}$ as the sum of the terms that enter
negatively in Eq. (\ref{Apsi}) and $v^{2}$ as the sum of the terms
that enter positively in Eq. (\ref{Apsi}). Obviously
$u$ and
$v$ are real and  positive. The relevant integration is
only over these two variables and is of the form
\begin{equation}
\int_{0}^{u_{\rm max}}\!du\int_{0}^{v_{\rm max}}\!dv
\;{h(u,v)\over \omega-u^{2}+v^{2}},\label{numeratorh}
\end{equation}
where
\begin{equation}
\omega=k_{0}-(n-n')E(k/[n-n']).\label{defomega}
\end{equation}
A branch point in the self-energy is now reduced to the question of
showing that Eq. (\ref{numeratorh}) has a branch point at $\omega=0$.
The limits $u_{\rm max}$ and $v_{\rm max}$ represent the region of
validity of the second order Taylor series expansion. The integral as
written has a branch cut on the real $\omega$ axis for $-v_{\rm
max}^{2}<\omega <u_{\rm max}^{2}$.

\subsection{Existence of the branch point}

The question at hand is whether the integral Eq. (\ref{numeratorh})
has, in addition to the branch cut along the real axis, a branch
point at
$\omega=0$.  The putative existence of such a branch point
clearly comes from the region $u\approx v$ and has nothing to do with the
upper limits of integration and nothing to do with the numerator
function $h(u,v)$. To complete the analysis it is therefore sufficient to
examine the function
\begin{equation}
f(\omega)=\int_{0}^{M}\!du\int_{0}^{M}\!dv\;{1\over\omega-u^{2}+v^{2}}
.\end{equation}
Although this integral cannot be performed explicitly, it is possible to
prove the existence of a branch point at $\omega=0$.
To analyze this integral it is useful to split the integration
over $v$ into two parts:
\begin{eqnarray}
f(\omega)=&&\int_{0}^{M}\!du\int_{0}^{u}\!dv\;{1\over
\omega-u^{2}+ v^{2}}\nonumber\\
+&&\int_{0}^{M}\!du\int_{u}^{M}\!dv\;{1\over
\omega-u^{2}+ v^{2}}.\nonumber
\end{eqnarray}
In the first integral, replace   $v$ by
$x=\sqrt{u^{2}-v^{2}}$. In the second, replace $v$ by
$x=\sqrt{v^{2}-u^{2}}$: 
\begin{eqnarray}
f(\omega)=&&\int_{0}^{M}\!du\int_{0}^{u}\! dx
\:{x\over\sqrt{u^{2}-x^{2}}}\;{1\over \omega-x^{2}}\nonumber\\
+&&\int_{0}^{M}\!du\int_{0}^{\sqrt{M^{2}-u^{2}}}\!\!dx
\;{x\over\sqrt{u^{2}+x^{2}}}\;{1\over \omega+x^{2}}.\nonumber
\end{eqnarray}
Now interchange the order of integration in both and perform
the integrations over $u$ to obtain
\begin{displaymath}
f(\omega)\!=\!\int_{0}^{M}\! dx\bigg({x\over\omega-x^{2}}
+{x\over\omega+x^{2}}\bigg)
\ln\!\bigg[{M+\sqrt{M^{2}-x^{2}}\over x}\bigg].
\end{displaymath} 
Note that  $f(0)$ is finite, but
$\big[df(\omega)/d\omega\big]_{\omega=0}$ is divergent as are all
the odd derivatives. This already shows that
$f(\omega)$ is not analytic at $\omega=0$. (Note that the original
function in Eq. (\ref{numeratorh}) contains a numerator $h(u,v)$
which could vanish at $x=0$. Consequently the divergence of the
first derivative of $f(\omega)$ may not hold when the numerator is
included. However,  higher derivatives will
diverge.)

To confirm the branch point at $\omega=0$ the best procedure is
to analytically continue $\omega$ in a small circle enclosing
the origin.  The integrand of $f(\omega)$ has simple poles at
$x_{1}=\sqrt{\omega}$, $x_{2}=-\sqrt{\omega}$,
$x_{3}=i\sqrt{\omega}$, and $x_{4}=-i\sqrt{\omega}$.
When $\omega$ has a small, positive imaginary part, these
singularities are off the real axis.  To expose the
branch point at
$\omega=0$, set $\omega=re^{i\phi}$. Then as $\phi$ increases from
$0^{+}$ to $2\pi^{+}$, all four $x_{j}$ move in small
counter-clockwise circles and return to different values:
$x_{1}$ moves to the negative real axis without coming near the
integration contour; $x_{2}$ moves counter-clockwise through the
integration contour into the upper half-plane at the position
originally occupied by $x_{1}$. The change in the value of the
integral can be computed by integrating in a small circular
contour $C_{1}$ around the position originally occupied by
$x_{1}$:
\begin{eqnarray}
\oint_{C_{1}}\!dx\,{x\over
\omega-x^{2}}\ln\!\bigg[{M+\sqrt{M^{2}-x^{2}}\over
x}\bigg]&&\nonumber\\
=i\pi \ln\!\bigg[{M+\sqrt{M^{2}-\omega}\over
\sqrt{\omega}}\bigg].&&\label{c1}
\end{eqnarray}
Likewise, as $\phi$ increases from
$0^{+}$ to $2\pi^{+}$, $x_{3}$ moves from the positive imaginary
axis clockwise to the negative imaginary axis without touching
to the real axis. However, $x_{4}$ moves counter clockwise from
the negative real axis to the positive real axis and drags the $x$
contour with it. The change in the value of the integral from this
distortion can be computed by integrating in a small circular
contour $C_{3}$ around the position originally occupied by $x_{3}$:
\begin{eqnarray}
\oint_{C_{3}}\!dx\,{x\over
\omega+x^{2}}\ln\!\bigg[{M+\sqrt{M^{2}-x^{2}}\over
x}\bigg]&&\nonumber\\
=-i\pi \ln\!\bigg[{M+\sqrt{M^{2}+\omega}\over
i\sqrt{\omega}}\bigg].&& \label{c2}
\end{eqnarray}
The change in $f(\omega)$ resulting from encircling the origin is
the sum of (\ref{c1}) and (\ref{c2}):
\begin{equation}
f\big(e^{2\pi i}\omega\big)\!-\!f(\omega)=
i\pi \ln\!\bigg[{i\big(M+\sqrt{M^{2}-\omega}\big)\over
M+\sqrt{M^{2}+\omega}}\bigg].
\end{equation}
The non-vanishing of the right hand side confirms that there is a
branch point at $\omega=0$ and completes the proof. 

The branch point  has infinitely many sheets because
if one rotates the phase of $\omega$ by $2\pi N$ the result is
\begin{equation}
f\big(e^{2\pi N i}\omega\big)\!-\!f(\omega)=
iN\pi \ln\!\bigg[{i\big(M+\sqrt{M^{2}-\omega}\big)\over
M+\sqrt{M^{2}+\omega}}\bigg].
\end{equation}

\section{Detailed proof of branch point at \lowercase{$k_{0}=\pm
k$}}

In Sec. III.C it was shown that for a real or complex 
single-particle energy with the asymptotic behavior in Eq.
(\ref{asympt}) the necessary conditions for an singularity are
satisfied  at $k_{0}=\pm k$.
This Appendix proves  that there  is a branch
point.

As before, it is necessary to examine the integration in the region
at which the contour is trapped.  
The denominator function is
\begin{equation}
\psi=\sum_{j=1}^{n}{\cal E}(\vec{p}_{j})-
\sum_{j=n+1}^{n+n'}{\cal E}(\vec{p}_{j})\label{Bpsi1}.
\end{equation}
Define the two momenta
\begin{eqnarray}\vec{s}={\hat{k}\over n}\big({k\over
2}+P_{\|}\big);\hskip1cm
\vec{s}^{\,\prime}={\hat{k}\over n'}\big({k\over
2}-P_{\|}\big),
\end{eqnarray}
and put
  \begin{equation}
 \vec{p}_{j}=\cases{\vec{s}+\vec{\alpha}_{j}\sqrt{P_{\|}},
   & $1\le j\le n$\cr
   \vec{s}^{\,\prime}+\vec{\alpha}_{j}\sqrt{P_{\|}}, & $n+1\le j\le
n+n'$.}
   \end{equation} 
Momentum conservation requires that
\begin{equation}
0=\sum_{j=1}^{n+n'}\vec{\alpha}_{j}.\label{Bmomentum}
\end{equation}
When all the $\vec{\alpha}_{j}=0$
\begin{displaymath}
\psi\big|_{\vec{\alpha}_{j}=0}=n{\cal E}(\vec{s})-n'{\cal
E}(\vec{s}').
\end{displaymath}
To demonstrate that there is a branch point it is necessary to
expand $\psi$ in a Taylor series for $|\vec{\alpha}_{j}|\sqrt{P_{\|}}$
small compared to
$|\vec{s}|$ and $|\vec{s}'|$:
\begin{eqnarray}
\psi=&&nE(s)+{P_{\|}\over 2}\sum_{j=1}^{n}\bigg(
A_{\|}\alpha_{j\|}^{2}+
A_{\bot}\vec{\alpha}_{j\bot}^{2}\bigg)\nonumber\\
-&&n'E(s')-{P_{\|}\over 2}\sum_{j=n+1}^{n+n'}\bigg(
A_{\|}^{\prime}\alpha_{j\|}^{2}+
A_{\bot}^{\prime}\vec{\alpha}_{j\bot}^{2}\bigg).\nonumber
\end{eqnarray}
The terms linear in $\vec{\alpha}_{j}$ canceled by momentum
conservation as in Appendix A. The coefficients $A_{\|}$ and
$A_{\bot}$ are as defined in Eq. (\ref{defA}).

The branch point we are seeking occurs when $P_{\|}\to\infty$. In this
limit both $s$ and
$s'$ approach infinity so that
\begin{eqnarray}
{\cal E}(s)\to && s+{m^{2}\over 2s}+\dots\nonumber\\
{d{\cal E}\over ds}\to && 1-{m^{2}\over 2s^{2}}+\dots\nonumber\\
A_{\|}\to && {m^{2}\over 3s^{2}}+\dots\nonumber\\
A_{\bot}\to && {1\over s}-{m^{2}\over 2s^{3}}+\dots\nonumber
\end{eqnarray}
At large $P_{\|}$ the denominator function $\psi$ behaves as
\begin{eqnarray}
\psi\to && k+{(mn)^{2}\over 2P_{\|}+k}-{(mn')^{2}\over 2P_{\|}-k}
\nonumber\\
+&&{nP_{\|}\over 2P_{\|}+k}\sum_{j=1}^{n}\vec{\alpha}_{j\bot}^{2}
-{n'p_{\|}\over 2P_{\|}-k}\sum_{j=n+1}^{n+n'}\vec{\alpha}_{j\bot}^{2}.
\nonumber\end{eqnarray}
Now  take $P_{\|}\to\infty$ so that
\begin{equation}
P_{\|}\to\infty:
\hskip0.5cm \psi=k+{n\over 2}\sum_{j=1}^{n}\vec{\alpha}_{j\bot}^{2}
-{n'\over 2}\sum_{j=n+1}^{n+n'}\vec{\alpha}_{j\bot}^{2}.
\end{equation}
This is of the same form as Eq. (\ref{Apsi2}) except that
$\vec{\alpha}_{j\|}$ do not enter. Therefore, the proof 
from Appendix A applies. 
The contribution to the retarded self-energy  is
\begin{equation}
\Pi_{R}(k_{0})=\int\Big(\prod_{j=1}^{n+n'}\!d^{3}\alpha_{j}
\Big)\;{f(\vec{\alpha}_{j})\over k_{0}-\psi}
\end{equation}
The proof in Appendix A shows that if $k_{0}=k+re^{i\phi}$ then
the value of $\Pi_{R}(k_{0})$ does not return to the same value when
$\phi$ increases from 0 to $2\pi$.

\section{$\phi^{3}$ Example to one loop}

The simplest example of an essential singularity at $k_{0}=\pm k$ occurs
in a theory with ${\cal H}_{I}=g\phi^{3}/3!$. With the free-particle
dispersion relation $E(p)=(p^{2}+m^{2})^{1/2}$, the one-loop self-energy
has a branch cut for
$-k\le k_{0}\le k$ that results from an intermediate state with
$n=n'=1$. The following calculation will expose the essential singularity
at
$k_{0}=\pm k$ in the self-energy
\begin{displaymath}
\Pi_{R}(k_{0})=g^{2}\int\!{d^{3}p\over (2\pi)^{3}}
{n(p)-n(\vec{p}+\vec{k})\over
\big(k_{0}+i\epsilon-\psi\big)2E(p)2E(\vec{p}+\vec{k})}. 
\end{displaymath}
The denominator function is
\begin{displaymath}
\psi=E(\vec{p}+\vec{k})-E(p).
\end{displaymath}
 Rather that calculate $\Pi_{R}$ itself, it is 
easier to calculate the imaginary part:
\begin{displaymath}
{\rm Im}\Pi_{R}(K)=-{g^{2}\pi\over 4}\int\! {d^{3}p\over
(2\pi)^{3}}\,\delta[\,k_{0}-\psi\,]{n(p)-n(\vec{p}\!+\!\vec{k})\over
E(p)E(\vec{p}\!+\!\vec{k})}.
\end{displaymath}
With the decomposition $\vec{p}=\hat{k}p_{\|}+\vec{p}_{\bot}$,
the integral over $\vec{p}_{\bot}$ can be performed using the Dirac
delta function:
\begin{displaymath}
\int\! d^{2}p_{\bot}\,\delta[k_{0}-\psi]={2\pi p_{\bot}\over 
d\psi/dp_{\bot}}\bigg|_{\psi=k_{0}}
=2\pi{E(\vec{p})E(\vec{p}\!+\!\vec{k})\over k_{0}}.
\end{displaymath}
Consequently
\begin{equation}
{\rm Im}\Pi_{R}(K)=-{g^{2}\over 16\pi k_{0}}\int_{p_{\|}^{\rm
min}} ^{\infty}\!dp_{\|}
\Big[n(p)-n(p+k)\Big].\label{impi}
\end{equation}
The condition $k_{0}=\psi$ can only be satisfied for $K^{2}<0$ and it  
makes $E(\vec{p})$ a linear function of
$p_{\|}$:
\begin{displaymath}
E(\vec{p})=
{k\over k_{0}}p_{\|}-{K^{2}\over 2k_{0}},\end{displaymath}
where
\begin{displaymath}
p_{\|}=
-{k\over 2}+{k_{0}\over 2}\sqrt{1-{4(m^{2}+p_{\bot}^{2})\over K^{2}}}.
\end{displaymath}
Since $0\le p_{\bot}\le \infty$,  the minimum value  of the parallel
momentum is 
\begin{displaymath}
p_{\|}^{\rm min}=-{k\over 2}+{k_{0}\over
2}\sqrt{1-{4m^{2}\over K^{2}}}.
\end{displaymath}
The remaining integration in (\ref{impi}) is elementary:
\begin{equation}
{\rm Im}\Pi_{R}(K)=-{g^{2}T\over 16\pi k}\ln\Bigg[{1-e^{-\beta
E(k+p_{\|}^{\rm min})}\over 1-e^{-\beta
E(p_{\|}^{\rm min})}}\Bigg],\label{impi2}
\end{equation}
where the energies that enter  are
\begin{eqnarray}
E(p_{\|}^{\rm min})=&&
-{k_{0}\over 2}+{k\over 2}\sqrt{1-{4m^{2}\over K^{2}}}\nonumber\\
E(k\!+\!p^{\rm min}_{\|})=&&
{k_{0}\over 2}+{k\over 2}\sqrt{1-{4m^{2}\over K^{2}}}.
\nonumber\end{eqnarray}
Both energies are positive since $K^{2}<0$.
Thus the imaginary part is
\begin{displaymath}
{\rm Im}\Pi_{R}(K)=-{g^{2}T\over 16\pi k}\ln\Bigg[{1-e^{-\beta
\big(k_{0}+k\sqrt{1-4m^{2}/K^{2}}\big)/2}\over 1-e^{-\beta
\big(-k_{0}+k\sqrt{1-4m^{2}/K^{2}}\big)/2}}\Bigg].
\end{displaymath}
The imaginary part is an odd function of $k_{0}$.
As expected, there is an essential singularity at $k_{0}=\pm k$.
The leading behavior as $k_{0}\to k$ is
\begin{eqnarray}
{\rm Im}\Pi_{R}(K)\to - &&{g^{2}T\over 16\pi k}\big( e^{\beta
k/2}-e^{-\beta k/2}\big)\nonumber\\
\times&&\exp\Bigg(\!-{\beta k\over 2}\sqrt{1-{2(nm)^{2}\over
k(k_{0}-k)}}\;\Bigg).\label{essential1}
\end{eqnarray}
This agrees perfectly with Eq. (\ref{statistical3}) for $n=n'=1$.

\section{$\phi^{4}$ example at two loops}

This appendix will explicitly show the  branch points on the light
cone and on the mass shell in $\phi^{4}$ theory at two-loop order.
The analysis of this section will be based on the work of Wang and
Heinz \cite{Heinz}, who calculated the imaginary part of the
self-energy to  two-loop order as a function of energy. Previous
works had  computed the imaginary part just on the mass-shell
\cite{Parwani,Jeon}.

The notation in this appendix will be that of Wang and Heinz
\cite{Heinz}. The interaction Hamiltonian is 
$g^{2}\phi^{4}/4!$. The zero-temperature particles are taken as
massless, but thermal resummation leads to propagators with poles
at
$p_{0}=\pm(p^{2}+m_{P}^{2})^{1/2}$, where
$m_{P}$ is a  resummed plasmon mass
\begin{equation}
m_{P}^{2}={g^{2}T^{2}\over 24}\bigg(1-{g\over 2\pi}\sqrt{{3\over
2}} \bigg).\end{equation}
The imaginary part of the two-loop self-energy is grouped as
\begin{displaymath}
{\rm Im}\Sigma(\omega,\vec{p})={\rm Im} g_{1}(\omega,\vec{p})
+{\rm Im} g_{2}(\omega,\vec{p}).
\end{displaymath}
Here $g_{1}$ contains the usual three-particle cuts (i.e. 
$n=3$, $n'=0$ and $n=0$, $n'=3$) and will not be discussed here;
$g_{2}$ contains the cut for two emissions and one absorption
($n=2$, $n'=1$) and the cut for one emission and two absorption
($n=1$, $n'=2$). As demonstrated in Section III, the function
$g_{2}(\omega,\vec{p})$ should have  branch points at
$\omega=\pm(p^{2}+m_{P}^{2})^{1/2}$ and  essential singularities at
$\omega=\pm p$. 
The results of Wang and Heinz for ${\rm Im}\,
g_{2}(\omega,\vec{p})$ are extremely complicated double integrals
when $\vec{p}\neq 0$. Consequently, this Appendix will only
examine $\vec{p}=0$ and will demonstrate a branch point at
$\omega=m_{P}$ and an  essential singularity at $\omega=0$.

At $\vec{p}=0$, Wang and Heinz express the imaginary part of the
self-energy as
\begin{eqnarray}
\omega<m_{P}:\hskip0.6cm  {\rm Im} g_{2}(\omega,0)= &&
\int_{\varepsilon}^{\infty}\!\! dv\, F(w, v)\label{less}\\
\omega>m_{P}:\hskip0.6cm {\rm Im}
g_{2}(\omega,0)= &&\int_{a}^{\infty}\!\! dv\, F(w, v)
.\label{more}\end{eqnarray}
In the integrals, $v$ and $w$ are Latin letters: $v$ is a
dimensionless  variable (an energy divided by T) and $w$ is the
dimensionless ratio
\begin{displaymath}
w={\omega\over T}.
\end{displaymath}
The lower limit of the first integral is
\begin{equation}
\varepsilon=\bigg[a^{2}+{(a^{2}-w^{2})
(9a^{2}-w^{2})\over 4w^{2}}\bigg]^{1/2},\label{varepsilon}
\end{equation}
and the lower limit  of the second integral is
\begin{displaymath}
 a={m_{P}\over T}.
\end{displaymath}
Note that $\varepsilon\to a$ when $w\to a$. 

\subsection{Branch point at $\omega=m_{P}$}

At the mass-shell $\omega=m_{P}$ (equivalently $w=a$) the lower
limits are equal ($\varepsilon=a$) and thus ${\rm
Im}g_{2}(\omega,0)$ is continuous at
$\omega=m_{P}.$ 

The first derivative of Eq. (\ref{less}) is
\begin{eqnarray}
\omega<m_{P}:\hskip0.3cm  T\,{d\,{\rm Im} g_{2}(\omega,0)
\over d\omega}= &&-{d\varepsilon\over
dw}\,F(w,\varepsilon)\nonumber\\ +&&\int_{\varepsilon}^{\infty}\!\!
dv\,{\partial F(w, v)
\over \partial w}.
\end{eqnarray}
An essential property of the integrand is \cite{Heinz}
\begin{equation}
F(w,v)\big|_{v=a}=0.\label{essential}
\end{equation}
As $\omega\to m_{P}$,  the lower limit
$\varepsilon\to a$.  Since $F(a,a)=0$ by Eq. (\ref{essential}),
the first derivative of Eq. (\ref{less})  
at $\omega=m_{P}$ is the same as the first derivative of Eq.
(\ref{more}) at $\omega=m_{P}$.

The second derivative of Eq. (\ref{less}) is
\begin{eqnarray}
\omega<m_{P}:\hskip0.3cm  T^{2}\,{d^{2}{\rm Im} g_{2}(\omega,0)
\over d\omega^{2}}= &&-{d\over dw}\bigg[{d\varepsilon\over
dw}\,F(w,\varepsilon)\bigg]
\nonumber\\ -{d\varepsilon\over dw}
\,{\partial F(w, \varepsilon)
\over \partial w}
+&&\int_{\varepsilon}^{\infty}\!\!
dv\,{\partial^{2} F(w, v)
\over \partial w^{2}}.
\end{eqnarray}
The second term on the right hand side vanishes at $w=a$
because of Eq. (\ref{essential}). The first term on
the right hand side simplifies to
\begin{displaymath}
-{d^{2}\varepsilon\over dw^{2}}\,F(w,\varepsilon)-{d\varepsilon
\over dw}\,{\partial F(w,\varepsilon)\over \partial w}
-\bigg[{d\varepsilon\over dw}\bigg]^{2}{\partial
F(w,\varepsilon)\over d\varepsilon}.
\end{displaymath}
The first two terms of this vanish at $w=a$ because
of Eq. (\ref{essential}), but the third does not. 
Thus as $\omega$ approaches $m_{P}$ from below, the second
derivative is
\begin{eqnarray}
T^{2}\,{d^{2}{\rm Im} g_{2}(\omega,0)
\over d\omega^{2}}\bigg|_{\omega=m_{P}^{-}}= &&-
\bigg[{d\varepsilon\over dw}\bigg]_{\omega =a}^{2}
{\partial F(a,v)\over\partial v}\Bigg|_{v=a}
\nonumber\\ \nonumber\\
&&+\int_{a}^{\infty}\!\!
dv\,{\partial^{2} F(w, v)
\over \partial w^{2}}.
\end{eqnarray}
At $\omega=a$, Eq. (\ref{varepsilon}) gives $d\varepsilon/dw=-2$.
From Wang and Heinz \cite{Heinz} the function $F(w,v)$ simplifies
at $w=a$ to
\begin{eqnarray}
F(a,v)={g^{4}T^{2}\over 128\pi^{3}}{e^{v}\over e^{v}-1}
{e^{a}-1\over e^{a+v}-1}
2\ln\bigg[{\sinh(v/2)\over\sinh(a/2)}\bigg].
\end{eqnarray}
As expected, this vanishes at the lower limit $v=a$.
However $\partial F(a,v)/\partial v$ does not vanish at $v=a$. 
Thus the second derivative is discontinuous at $w=m_{P}$ with a
discontinuity given by
\begin{eqnarray}
\Bigg[{d^{2}{\rm Im} g_{2}(\omega,0)
\over d\omega^{2}}\Bigg]_{\omega=m_{P}^{-}}^{\omega=m_{P}^{+}}
={g^{4}\over 32\pi^{3}}{e^{a}\over
(e^{a}-1)^{2}}.\label{2derivative}
\end{eqnarray}
This confirms the existence of a branch point at the mass-shell
$\omega=m_{P}$.

There is a further check of Eq. (\ref{2derivative}). 
In the specific calculation of Wang and Heinz, the mass was
entirely thermal so that $a=m_{P}/T$ is independent of temperature.
However, the calculation of the two-loop discontinuity would also
apply in a theory with a non-thermal mass $m$. Then the right hand
side of Eq. (\ref{2derivative}) would be temperature-dependent with
$a=m/T$. In the zero-temperature limit, $a\to\infty$ and
discontinuity in the second derivative vanishes as expected.

\subsection{Essential singularity at $\omega=0$}

For $\vec{p}\neq 0$, the function $g_{2}(\omega,\vec{p})$ will have
essential singularities at $\omega=\pm p$. In the case considered
here, viz. $\vec{p}=0$, these collapse to an essential singularity
at $\omega=0$. 
In the vicinity of $\omega\approx 0$, the imaginary part is given
by Eq. (\ref{less}). The lower limit of the integral grows as
$\omega\to 0$:
\begin{displaymath}
\omega\to 0:\hskip0.75cm \varepsilon\to {3a^{2}\over
2w}+\dots
\end{displaymath}
It is convenient to change variables from $v$ to $\overline{v}$,
\begin{equation}
v={3a^{2}\over 2w}\big[1+\overline{v}\big],
\end{equation}
where $0\le\overline{v}\le\infty$. Then Eq. (\ref{less}) becomes
\begin{equation}
\omega< m_{P}:\hskip0.5cm {\rm Im} g_{2}(\omega,0)={3a^{2}\over
2w}\int_{0}^{\infty}
\!d\overline{v}\;F(w,v).
\end{equation}
From \cite{Heinz} the integral becomes in the limit $\omega\to
0$:
\begin{eqnarray}
&&{\rm Im} g_{2}(\omega,0)\to {9g^{4}T^{2}\over 1024 \pi^{3}}
\,{a^{4}\over w}\,e^{-3a^{2}/2w}\;\; I(\omega)\\
\nonumber\\
&&I(\omega)=
\int_{0}^{\infty}\!d\overline{v}\,e^{-3a^{2}\overline{v}
/2\omega}\big(1+\overline{v}\big)\Big(1+\sqrt{{3\overline{v}\over
4+3\overline{v}}}
\Big).\nonumber
\end{eqnarray}
As $w\to 0$, the integrand of $I(\omega)$ is exponentially small
for any  $\overline{v}$ that is not infinitesimal. The
dominant contribution comes from the region $0\le\overline{v}\ll
2w/3a^{2}$ and gives $I(\omega)\to 2\omega/3a^{2}$. 
Thus 
\begin{equation}
\omega\to 0:\hskip0.7cm{\rm Im} g_{2}(\omega,0)\to
{3g^{4}m_{P}^{2}\over 512
\pi^{3}}
\;e^{-3a^{2}/2w}.\label{essential2}
\end{equation}
The exponent here, $-3a^{2}/2w$, agrees precisely with that
anticipated in Eq. (\ref{statistical4}) for $n=2, n'=1$.

\end{document}